\documentstyle[aps,12pt,tighten,floats,epsf]{revtex}

\newcommand{\ldt}{${\cal L}$}
\newcommand{\nlsp}{$\tilde\chi^0_1$}
\newcommand{\tlsp}{$\tilde\tau_1$}

\newcommand{\cn}{$\tilde\chi^\pm_1\tilde\chi^0_2$}
\newcommand{\cc}{$\tilde\chi^\pm_1\tilde\chi^\pm_1$}

\newcommand{\met}{$\rlap{\kern0.25em/}E_T$}
\newcommand{\gmet}{$\gamma\rlap{\kern0.25em/}E_T$}
\newcommand{\ggmet}{$\gamma\gamma\rlap{\kern0.25em/}E_T$}
\newcommand{\gjjmet}{$\gamma jj \rlap{\kern0.25em/}E_T$}
\newcommand{\gbjmet}{$\gamma bj \rlap{\kern0.25em/}E_T$}
\newcommand{\gbbmet}{$\gamma b\bar{b} \rlap{\kern0.25em/}E_T$}
\newcommand{\gdjjmet}{$\gamma^\prime jj \rlap{\kern0.25em/}E_T$}
\newcommand{\lljjmet}{$\ell^\pm\ell^\pm jj \rlap{\kern0.25em/}E_T$}
\newcommand{\gmetjj}{$\gamma\rlap{\kern0.25em/}E_T+\ge 2\ {\rm jets}$}
\newcommand{\llmet}{$\ell\ell\rlap{\kern0.25em/}E_T$}
\newcommand{\lldedx}{$\ell\ell+dE/dx$}
\newcommand{\lllj}{$\ell\ell\ell j\rlap{\kern0.25em/}E_T$}
\newcommand{\lljj}{$\ell^\pm\ell^\pm jj\rlap{\kern0.25em/}E_T$}
\newcommand{\rsb}{$N_s/\delta N_b$}
\newcommand{\lsim}
    {\mathrel{\raise.3ex\hbox{$<$\kern-.75em\lower1ex\hbox{$\sim$}}}}
\newcommand{\gsim}
    {\mathrel{\raise.3ex\hbox{$>$\kern-.75em\lower1ex\hbox{$\sim$}}}}

\begin{document}


\vspace*{2.0cm}
\begin{center} 
{\large\bf Sensitivity to Gauge-Mediated Supersymmetry Breaking Models
       \\ \vspace*{0.3cm} of the Fermilab Upgraded Tevatron Collider}
  
   \vspace*{3.0cm}
   \large
   Jianming Qian \\ \vspace*{0.4cm}
   {\it Department of Physics, The University of Michigan \\
   Ann Arbor, Michigan 48109 \\
   Email: qianj@umich.edu} \\
\end{center}

\vspace*{3.0cm}

\begin{abstract}
   This paper discusses supersymmetry discovery potential of the upgraded 
   D\O\ experiment at the Tevatron $p\bar{p}$ collider. Six final states 
   with large transverse energy (momentum) leptons or photons (with or
   without large transverse momentum imbalances) are studied. These final 
   states are expected to have small backgrounds and are thereby ideal for 
   new physics searches. Implications of the analyses of these final states 
   on Gauge Mediated Supersymmetry Breaking models are discussed 
   for integrated luminosities of 2 and 30~fb$^{-1}$. This study 
   demonstrates that a large class of supersymmetry models can be 
   probed at the upgraded Tevatron.

\end{abstract}

\newpage
\section{Introduction}

This paper summarizes the work done for the Tevatron Run II 
Higgs/Supersymmetry workshop~\cite{workshop} on the supersymmetry
models with Gauge Mediation/Low Scale Supersymmetry 
Breaking~(GMSB)~\cite{gmsb}. Six final states in which new physics might 
manifest itself are investigated using the parameters of the upgraded D\O\ 
detector~\cite{d0det}. All of these final states are expected to have small 
physics and instrumentation backgrounds. Implications of the analyses of 
these final states in future Tevatron runs on the minimal (and not-so-minimal) 
GMSB models are discussed. Estimated discovery reaches in the supersymmetry 
parameter space are presented.

\section{Object Identification}

Due to a large number of Monte Carlo~(MC) events generated, no detector 
simulation is done for supersymmetry signals. All studies described in 
this paper except those extrapolated from Run~I analyses are carried out at 
the particle level of the {\sc Isajet} MC program~\cite{isajet}. A 2~TeV 
Tevatron center-of-mass energy is assumed throughout the studies.
Leptons ($\ell=e,\mu$) and photons ($\gamma$) are `reconstructed' from the 
generated particle list by requiring them to have transverse energy ($E_T$) or 
momentum ($p_T$) greater than 5~GeV and to be within the pseudorapidity 
ranges:
\begin{itemize}
  \item[] $e$:      $|\eta|<1.1$ or $1.5<|\eta|<2.0$;
  \item[] $\mu$:    $|\eta|<1.7$;
  \item[] $\gamma$: $|\eta|<1.1$ or $1.5<|\eta|<2.0$.
\end{itemize}
These fiducial ranges are dictated by the coverages of the electromagnetic
calorimeter and the central tracker of the D\O\ detector. Furthermore, 
the leptons and photons must be isolated. Additional energy in a cone with 
a radius ${\cal R}\equiv\sqrt{(\Delta\phi)^2+(\Delta\eta)^2}=0.5$ in
$\eta-\phi$ space around the lepton/photon is required to be 
less than 20\% of its energy.

Jets are reconstructed using a cone algorithm with a radius ${\cal R}=0.5$
in $\eta-\phi$ space and are required to have $E^j_T>20$~GeV and
$|\eta^j|<2.0$. All particles except neutrinos, the lightest supersymmetric
particles ({\sc lsp}), and the identified leptons and photons are used in 
the jet reconstruction. The transverse momentum imbalance (\met) is defined
to be the total transverse energy of neutrinos and the {\sc lsp}s.

Energies or momenta of leptons, photons and jets of Monte Carlo events are
taken from their particle level values without any detector effect. Smearing
of energies or momenta of leptons, photons and jets according to their 
expected resolution typically changes signal efficiencies by less than 10\%
relatively and therefore has negligible effect on the study.

The reconstruction efficiencies are assumed to be 90\% for leptons and photons.
For the purpose of background estimations, the probability ($\cal P$) for 
a jet to be misidentified as a lepton ($j\to\ell$) or a photon ($j\to\gamma$) 
is assumed to be $10^{-4}$. The probability for an electron to be 
misidentified as a photon ($e\to\gamma$) is assumed to be $4\times 10^{-4}$, 
These probabilities are a factor of three or more smaller than those obtained 
in Run~I~\footnote{The typical numbers determined in Run I are 
${\cal P}(j\to e)=5\times 10^{-4}$, ${\cal P}(j\to\gamma)= 7\times 10^{-4}$,
and ${\cal P}(e\to\gamma) = 4\times 10^{-3}$.}. With a new magnetic central 
tracking system and a fine-segmented
preshower detector, they should be achievable in Run~II.

In Run~I, tagging of b-jets was limited to the use of soft muons in D\O. 
Secondary vertex tagging of b-jets will be a powerful addition in Run~II. 
For the studies described below, a tagging efficiency of 60\% is assumed 
for those b-jets with $E_T>20$~GeV and $|\eta|<2.0$. The probability 
${\cal P}(j\to b)$ for a light-quark or gluon jet to be tagged as a b-jet is 
assumed to be $10^{-3}$. These numbers are optimistic extrapolations of 
what CDF achieved in Run~I.

Heavy stable charged particles can be identified~\cite{gll,dedx} using their 
expected large ionization energy losses ($dE/dx$) in the silicon detector, 
fiber tracker, preshower detectors and calorimeter. Based on Ref.~\cite{gll}, 
a generic $dE/dx$ cut is introduced with an efficiency of 
68\% for heavy stable charged particles and a rejection factor of 10
for the minimum ionization particles~(MIP). Note that the 
efficiency for identifying at least one such particle in events with two 
heavy stable charged particles is 90\%.

With the addition of preshower detectors, D\O\ will be able
to reconstruct the distance of the closest approach~({\sc dca}) of a photon
with a resolution $\sim 1.5$~cm~\cite{gll}. Here the {\sc dca} is defined as 
the distance between the primary event vertex and the reconstructed 
photon direction. Thereby it will enable us
to identify photons produced at secondary vertices. In the following, a photon
is called displaced if its {\sc dca} is greater than 5.0~cm and is
denoted by $\gamma^\prime$. We further assume that the probability for a 
photon produced at the primary vertex to have the measured {\sc dca}$>5$~cm is 
${\cal P}(\gamma\to\gamma^\prime)=2\times 10^{-3}$ (about $3\sigma$).

All final states studied have large $E_T$ ($p_T$) leptons/photons with or 
without large \met.
Triggering on these events are not expected to have any problem. Nevertheless,
we assume a 90\% trigger efficiency for all the final states.

\newpage
\section{Final States}

Signatures for supersymmetry vary dramatically from one model to another.
They can also be very different for different regions of the parameter space
in a model. Furthermore, these signatures are generally not unique to 
supersymmetry. In fact, some of the signatures are also expected from other 
theories beyond the Standard Model. Instead of chasing after theoretical 
models (all of which except perhaps one are wrong anyway) a set of final 
states which are somewhat generic to many new physics models, including 
supersymmetric models, is identified. All of these final states are 
characterized by high $E_T$($p_T$) isolated leptons/photons with or without 
large missing transverse momentum and
are thus expected to have small physics and instrumental backgrounds.
In the following, we discuss selection criteria and estimate observable 
background cross sections for six such final states: 
\begin{table}[htbp]
  \begin{tabular}{llll}
    & \ggmet & di-photon events with large \met & \\
    & \gbjmet  & single-photon events with b-jets and large \met & \\
    & \gdjjmet & single displaced photon events with jets and large \met & \\
    & \lldedx & high $p_T$ di-lepton events with large ionization energy & \\
   \hspace*{1.0cm} & \lllj & tri-lepton events with jets and \met & \\
    & \lljj  & like-sign di-lepton events with jets and \met & \hspace*{1.0cm}
 \\
  \end{tabular}
\end{table}

\subsection{\ggmet\ Final State}
\label{sec:ggmet}

The D\O\ Collaboration reported a search~\cite{ggmet} for di-photon events 
with large \met~(\ggmet\ events) motivated by supersymmetric models with 
a light gravitino ($\tilde G$) as the {\sc lsp} from a data sample of an 
integrated luminosity (\ldt) of $106.3\pm 5.6$~pb$^{-1}$ in Run~I. 
The \ggmet\ events were selected by requiring two identified photons, 
one with $E_T^\gamma>20$~GeV and the other with $E_T^\gamma>12$~GeV, 
each within pseudorapidity $|\eta^{\gamma}|<1.1$ or $1.5<|\eta^{\gamma}|<2.0$, 
and a $\rlap{\kern0.25em/}E_T$ greater than 25~GeV. Two events satisfied all 
requirements.

\begin{figure}[htbp]
  \centerline{\epsfysize=3.5in\epsfbox{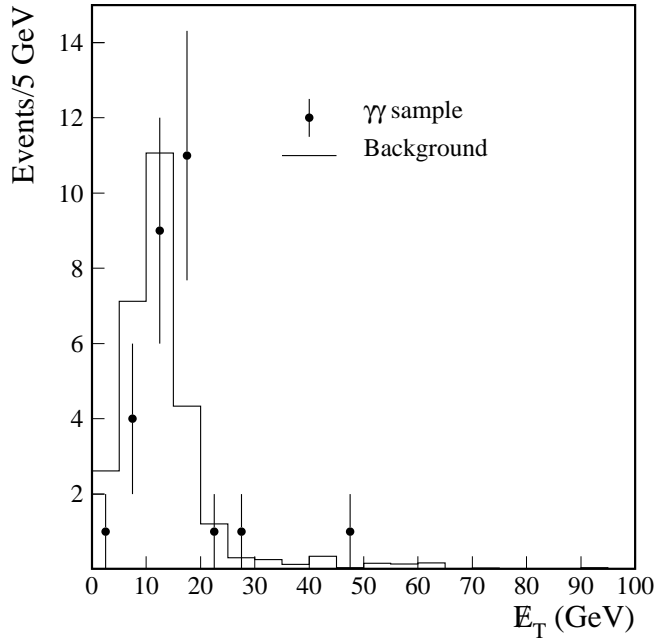}}
  \caption{The \protect$\rlap{\kern0.25em/}E_T$\ distributions of the 
           $\gamma\gamma$ and background samples for Run I. The number of 
           events with \protect$\rlap{\kern0.25em/}E_T$$<20$~GeV in the 
           background sample is normalized to that in the $\gamma\gamma$ 
           sample. Note that there is a 14~GeV \protect\met\ requirement
           in the trigger. The \protect\met\ values plotted here are 
           calculated off-line and therefore may differ from their values
           at the trigger level.}
  \label{fig:run1_ggmet_met}
\end{figure}

The principal backgrounds were multijet, direct photon, $W+\gamma$, 
$W+{\rm jets}$, $Z\rightarrow ee$, and $Z\rightarrow\tau\tau\rightarrow ee$
events from Standard Model processes with misidentified photons and/or 
mismeasured $\rlap{\kern0.25em/}E_T$. The numbers of estimated background 
events were $2.1\pm 0.9$ from \met\ mismeasurement~(QCD) and $0.2\pm 0.1$ from
misidentified photons~(fakes). This led to an observed background cross 
section of 20~fb from QCD and of 2~fb from fakes in Run~I. The \met\ 
distributions before the \met\ cut for both candidates and background events
are shown in Fig.~\ref{fig:run1_ggmet_met}. Note that events with large \met\
are rare.

Since the backgrounds are dominated by the \met\ mismeasurement, they can
be significantly reduced by raising the \met\ cut. Therefore, the following
selection criteria are used for the Run II studies:
\begin{itemize}
  \item[1)] At least two photons with $E^\gamma_T > 20$ GeV;
  \item[2)] \met$>50$ GeV.
\end{itemize}
The backgrounds with this set of selection criteria are expected to be 
significantly reduced by the increased cutoffs on \met\ and photon $E_T$ 
and by the improved photon identification. The total observable background 
cross section in Run II 
is estimated to be $\sigma_b = 0.4 ({\rm QCD}) + 0.2 ({\rm fakes})=0.6$~fb 
assuming reduction factors of 5 from the raised \met\ cutoff, 4 from the
improved ${\cal P}(j\to\gamma)$, 3 from the
higher photon $E_T$ requirement, and 10 from the decreased
$e\to\gamma$ fake probability.

\subsection{\gbjmet\ Final State}
\label{sec:gbjmet}

D\O\ Collaboration carried out a search~\cite{gjjmet} for single-photon 
events with at least two jets and large \met~(\gjjmet\ events) in Run~I. 
The \gjjmet\ events were selected by requiring at least one identified photon 
with $E_T^\gamma>20$~GeV and within pseudorapidity ranges $|\eta^\gamma|<1.1$ 
or $1.5<|\eta^\gamma|<2.0$, two or more jets having $E^j_T>20$~GeV and 
$|\eta^j|<2.0$, and $\rlap{\kern0.25em/}E_T>25$~GeV. A total of 318 events 
were selected from a data sample corresponding to an integrated luminosity of 
$99.4\pm 5.4$~pb$^{-1}$.

\begin{figure}[htbp]
  \centerline{\epsfysize=3.5in\epsfbox{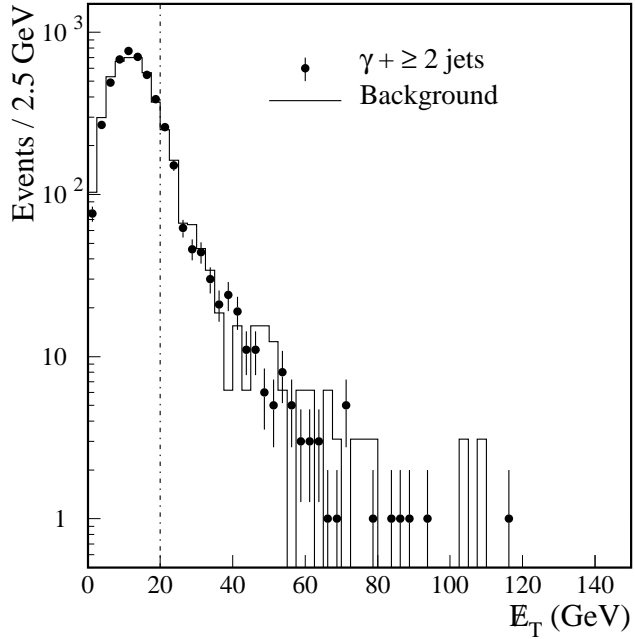}}
  \caption{The \protect\met\ distributions of the $\gamma jj$ and background 
           events. The number of events in the background is normalized to the 
           $\gamma jj$ sample for $\rlap{\kern0.25em/}E_T < 20$~GeV, the region 
           left of the dot-dashed line.}
  \label{fig:run1_gjjmet_met}
\end{figure}

The principal backgrounds were found to be QCD direct photon and multijet 
events, where there was mismeasured \met\ and a real or fake photon. 
The number of events from this source was estimated to be $315\pm 30$.
Other backgrounds such as those from $W$ with electrons misidentified as 
photons were found to be small, contributing $5\pm 1$ events. This led to 
an observed background cross section of 3,200~fb from the \met\ mismeasurement 
and of 50~fb from the fakes. The \met\ distribution before the \met$>25$~GeV 
cut is shown in Fig.~\ref{fig:run1_gjjmet_met}. As shown in the figure, 
the backgrounds can be significantly reduced by raising the requirement 
on \met. 

Events with a high $E_T$ photon, b-jets
and large \met\ are expected in some new physics models. These events,
referred to as \gbjmet, are in many ways similar to the \gjjmet\ events and 
thereby can be selected similarly:
\begin{itemize}
  \item[1)] At least one photon with $E^\gamma_T >20$~GeV;
  \item[2)] At least two jets with $E^j_T>20$~GeV; 
  \item[3)] At least one jet is tagged as a b-quark jet with $E^b_T>20$~GeV;
  \item[4)] \met$>50$~GeV;
  \item[5)] No leptons with $E^\ell_T>20$~GeV.
\end{itemize}
The backgrounds from the QCD multijet events with real or misidentified 
photons and from the W events with electrons faking photons are estimated to 
be 0.63~fb, assuming background reduction factors of 5 from the raised \met\ 
requirement, 2 from the improved photon identification and using the assumed 
value of ${\cal P}(j\to b)$. 
The dominant background sources are expected to be $\gamma b\bar{b}$ and
$\gamma t\bar{t}$ events. These background sources cannot be reduced by
the tagging of b-jets. However, the $\gamma b\bar{b}$ contribution is
expected to be small due to the large \met\ requirement. Monte Carlo studies 
show that it is negligible. The $\gamma t\bar{t}$ 
(with $t\bar{t}\to W^+W^-b\bar{b}$) contribution is reduced by the requirements 
4) and 5) and is estimated using the cross section of Ref.~\cite{gtt}. 
A total of 0.9~fb observable background cross section is assumed.

\subsection{\gdjjmet\ Final State}
\label{sec:gdjjmet}
Photons produced at secondary vertices are predicted in a class of
new physics models. These photons will appear to have large values of 
{\sc dca}. Though dramatic, they alone are unlikely to be 
sufficient to reduce backgrounds from cosmic rays or from mismeasurement.
We therefore select events with displaced photons accompanied by jets and 
large \met:
\begin{itemize}
  \item[1)] At least one displaced photon with $E^{\gamma^\prime}_T >20$~GeV;
  \item[2)] At least two jets with $E^j_T>20$~GeV;
  \item[3)] \met$>50$~GeV.
\end{itemize}
These are called \gdjjmet\ events. The dominant backgrounds are the same as
those for the \gjjmet\ events, with a vertex-pointing photon being 
misidentified as a displaced photon. Using ${\cal P}(\gamma\to\gamma^\prime)$,
the observable background cross section from QCD and W events is estimated to 
be 0.6~fb.

\subsection{High $p_T$ \lldedx\ Final State}
\label{sec:lldedx}
One possible new physics signature is the presence of heavy stable charged 
particles. These particles, if produced, will manifest themself in the 
detector as slowly moving muons with
large ionization energy losses. Though D\O\ had several di-lepton 
analyses in Run~I, none of these can be extrapolated to Run~II, 
thanks to the replacement
of the central tracker. Based on the expected signatures of several
supersymmetric models with heavy stable charged particles discussed below, 
we select high $p_T$ di-lepton events (\lldedx) with large $dE/dx$ loss 
using the following requirements:
\begin{itemize}
  \item[1)] At least two leptons with $p^\ell_T>50$~GeV;
  \item[2)] $M_{\ell\ell}>150$ GeV;
  \item[3)] At least one lepton passing the $dE/dx$ requirement.
\end{itemize}
The di-lepton mass requirement is intended to reduce Drell-Yan backgrounds.
The principal backgrounds are: QCD dijet events with jets misidentified as
leptons, $t\bar{t}$, and Drell-Yan events. Using ${\cal P}(j\to\ell)=10^{-4}$ 
and the assumed rejection factor of the $dE/dx$ cut for the MIP particles, 
the observable background cross sections are estimated to 
be 0.1~fb from QCD dijet, 0.2~fb from $t\bar{t}$ events, 
and 0.2~fb from Drell-Yan processes. The QCD dijet cross section for 
$p_T>50$~GeV is assumed to be 1~$\mu$b in the estimation. 
The total observable cross section is therefore 0.5~fb for the above selection.

\subsection{\lllj\ Final State}
\label{sec:lllj}

D\O\ searched for gaugino pair production using the tri-lepton 
signature~\cite{lllj} in Run~I. The lepton $p_T$ cut was typically
15~GeV for the leading lepton and 5~GeV for the non-leading leptons. 
The analysis also had a small \met\ requirement. The observable background
cross section was estimated to be around 13~fb. Most of these backgrounds are
due to Drell-Yan processes. We select the \lllj\ events using the following
criteria:
\begin{itemize}
  \item[1)] $p^{\ell_1}_T>15$~GeV, $p^{\ell_2}_T>5$~GeV, $p^{\ell_3}_T>5$~GeV;
  \item[2)] \met$>20$~GeV;
  \item[3)] At least one jet with $E^j_T>20$~GeV.
\end{itemize}
The Drell-Yan production, a major background source for the Run~I analysis,
is significantly reduced by the new jet requirement. The total observable
background cross section is estimated to be 0.3~fb assuming background
reduction factors of 10 from the jet requirement, 2 from the improved particle 
identification, 2 from the higher \met\ cut.

\subsection{\lljjmet\ Final State}
\label{sec:lljj}
Like-sign di-lepton events are expected from processes such as gluino pair
production. They are also expected from processes with three or more
leptons in the final states, but only two are identified. This 
final state is expected to have small backgrounds. Again without a magnetic 
tracker, D\O\ had no analysis of this nature in Run~I.
Based on Monte Carlo studies for several supersymmetric models, we select
\lljjmet\ events using the following criteria:
\begin{itemize}
  \item[1)] Two like-sign leptons  with $p^\ell_T>15$~GeV;
  \item[2)] At least two jets with $E^j_T>20$~GeV;
  \item[3)] \met$>25$~GeV.
\end{itemize}
Events with three or more identified leptons are removed to make the sample
orthogonal to the \lllj\ sample. Since leptons are relatively soft in 
$p_T$ for the new physics model we investigated using this selection,
the effect of charge confusion due to a limited tracking resolution
is thus neglected in this study.
The major backgrounds are: $W+{\rm jets}$ events with one of the jets 
misidentified as a lepton, $t\bar{t}$ events with energetic leptons from 
b-quark decays, and Drell-Yan ($WZ$, $ZZ$) events. The $W+{\rm jets}$ 
background is estimated using the number of $W+ 3j$ events observed in 
Run~I folded with ${\cal P}(j\to\ell)$ to be 0.2~fb. The $t\bar{t}$ and
Drell-Yan backgrounds are estimated using Monte Carlo to be 0.1 and 0.1~fb 
respectively. Adding the three background sources together yields
a total observable background cross section of 0.4~fb.

\newpage
\section{Constraints on GMSB Models}
The supersymmetric models with gauge mediated supersymmetry breaking are
characterized by a supersymmetry breaking scale $\Lambda$ as low as 100~TeV 
and a light gravitino which is naturally the lightest supersymmetric
particle. In these models, supersymmetry is assumed to be broken
in a hidden sector and the symmetry breaking is transmitted to the visible
sector of Standard Model particles and their superpartners through the 
Standard Model gauge interactions. The minimum gauge mediated supersymmetry
breaking model is described by five parameters:
\begin{table}[htpb]
  \begin{tabular}{llll}
    & $\Lambda$: & supersymmetry breaking scale  & \\
    \hspace*{2.0cm} & $M_m$: & messenger sector scale & \hspace*{2.0cm} \\
    & $N$:       & number of messengers &      \\  
    & $\tan\beta$: & ratio of the v.e.v of the two higgs doublets & \\
    & ${\rm sign}(\mu)$: &  sign of the higgs mass parameter & \\
  \end{tabular}
\end{table}

\begin{figure}[htbp]
  \centerline{\epsfysize=3.5in\epsfbox{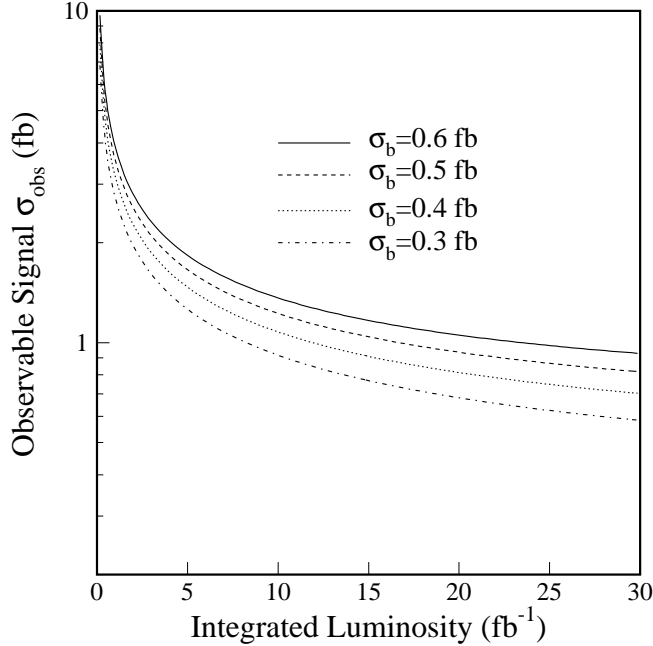}}
  \caption{The minimum observable signal cross section $\sigma_{obs}$ for 
           a $5\sigma$ discovery as a function of integrated luminosity
           for four different values of background cross sections.} 
  \label{fig:obs}
\end{figure}

The phenomenology is largely determined by the next-to-lightest supersymmetric
particle which in turn depends on the values of the above five
parameters. For a review of GMSB models, see Ref.~\cite{gmsb}. 
In the following, we discuss expected sensitivities with
integrated luminosities of 2, 30~fb$^{-1}$ for four different 
model lines defined by the working group. Each model line has a different
{\sc nlsp}. All theoretical expectations and signal efficiencies are obtained
from the {\sc Isajet} MC program. A minimum $p_T$ of 50~GeV of the hard 
scattering is applied for all 
signal processes. We define the significance ($N_s/\delta N_b$) as the ratio 
between the number ($N_s$) of expected signal events and the error 
($\delta N_b$) on the estimated number of background events. 
Here a 20\% systematic uncertainty is assumed for all estimated 
observable background cross sections. Therefore,
$$\delta N_b=\sqrt{{\cal L}\cdot\sigma_b + (0.2\cdot{\cal L}\cdot\sigma_b)^2}$$
We characterize the sensitivity using the minimum signal cross section
$\sigma_{dis}$ for a 5 standard deviation ($5\sigma$) discovery:  
$$\frac{N_s}{\delta N_b} = \frac{{\cal L}\cdot\sigma_{dis}\cdot\epsilon}
                                {\delta N_b} = 5$$
where $\epsilon$ is the efficiency for the signal. The minimum observable
signal cross section $\sigma_{obs}$ defined as $\sigma_{dis}\cdot\epsilon$
for the discovery is therefore independent of signal processes. 
The $\sigma_{obs}$ as a function of \ldt\ for several different values of 
$\sigma_b$ are shown in Fig.~\ref{fig:obs}. It decreases dramatically as
\ldt\ increases for small \ldt\ values and flattens out for large \ldt\
values. Clearly, the sensitivity can be improved for large \ldt\ values by
tightening the cuts to reduce backgrounds further.

In the following, we express the $5\sigma$ discovery cross sections as 
functions of the supersymmetry breaking scale $\Lambda$ and the lighter 
chargino ($\tilde\chi^\pm_1$) mass for the four different model lines.

\subsection{Model Line 1: $\tilde\chi^0_1$ as the NLSP}
\begin{figure}[htbp]
  \centerline{\epsfysize=3.5in\epsfbox{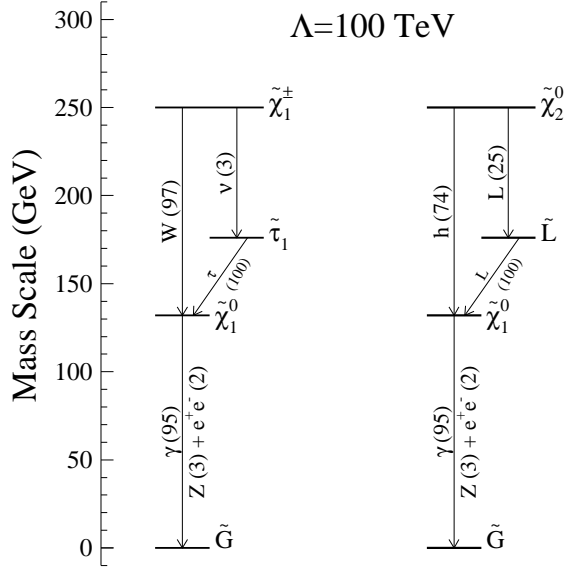}}
  \caption{Decay schematics of $\tilde\chi^\pm_1$ and $\tilde\chi^0_2$
           for $\Lambda=100$~TeV for the model line with a \protect\nlsp\ as
           the {\sc nlsp}. Percentage branching ratios for main decay modes
           are shown in parentheses.}
  \label{fig:p1br}
\end{figure}

Within the framework of the minimal GMSB models, \nlsp\ is the {\sc nlsp} for
most of the parameter space. If the \nlsp\ has a non-zero photino
component, it is unstable and decays to a photon plus a gravitino 
($\tilde\chi^0_1\to\gamma\tilde G$) with a branching ratio of nearly 100\%. 
Depending on its lifetime, pair production of supersymmetric particles 
will result in \ggmet, \gmet, and \met$+X$ events. For the purpose of this 
study, we consider a class of models with the following parameters fixed:
$$N=1,\ \ \frac{M_m}{\Lambda}=2,\ \ \tan\beta=2.5,\ \ \mu>0$$
while $\Lambda$ is allowed to vary. For the range of $\Lambda$ values 
of interest at the Tevatron, the supersymmetry production cross section is 
dominated by \cc\ and \cn\ production. Figure~\ref{fig:p1br} 
shows the schematic decay chains of $\tilde\chi^\pm_1$ and $\tilde\chi^0_2$ 
with their branching ratios for $\Lambda=100$~TeV. In the following, 
scenarios with prompt and delayed $\tilde\chi^0_1$ decays are discussed. 
If the $\tilde\chi^0_1$ is quasi-stable, {\it i.e.} with a long lifetime, 
the signature will be identical to that of the supersymmetric models with 
gravity mediation.

\subsubsection{Prompt $\tilde\chi^0_1\to\gamma\tilde G$ Decay}
\begin{figure}[htbp]
  \centerline{\epsfysize=3.0in\epsfbox{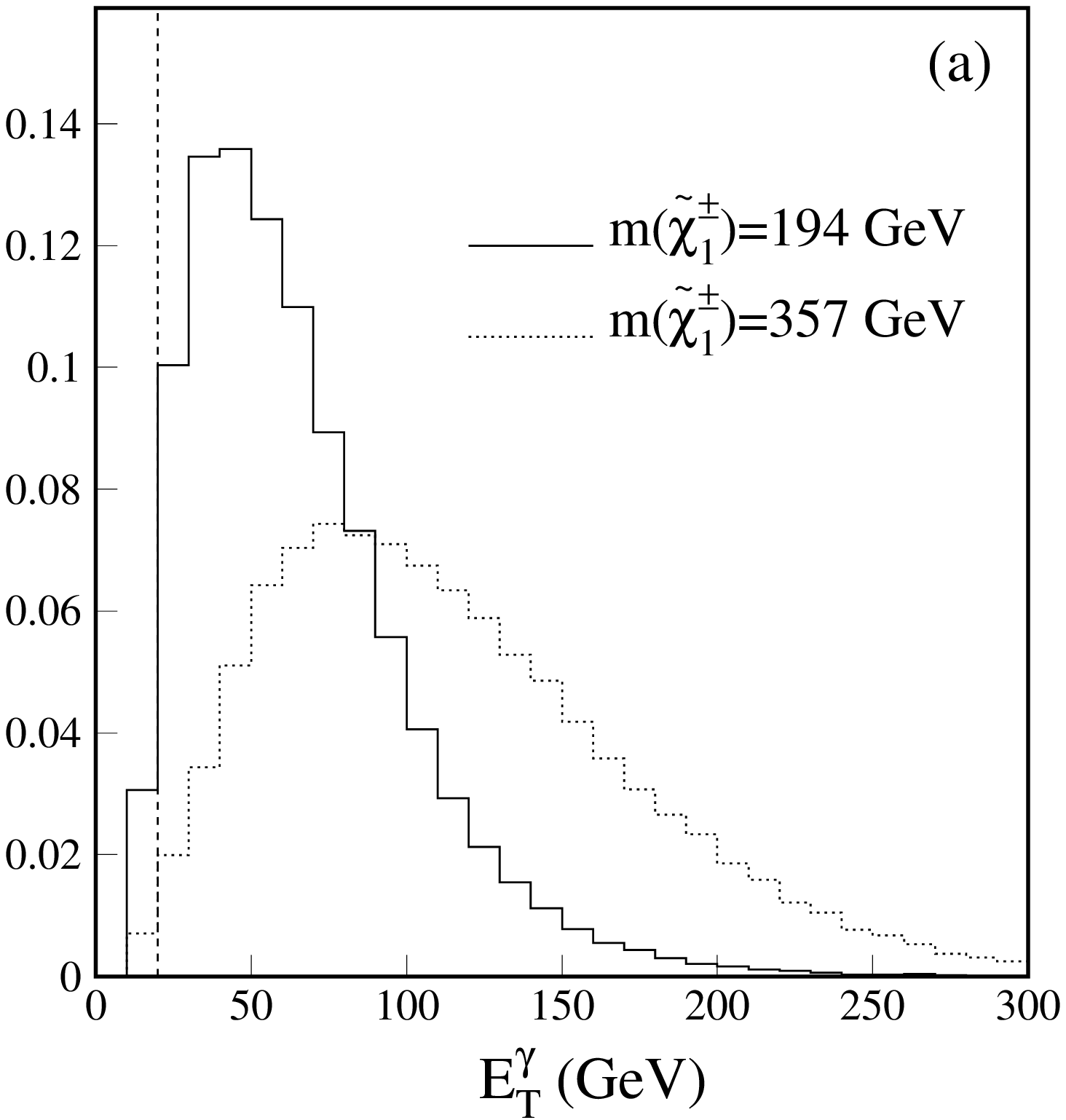}
              \epsfysize=3.0in\epsfbox{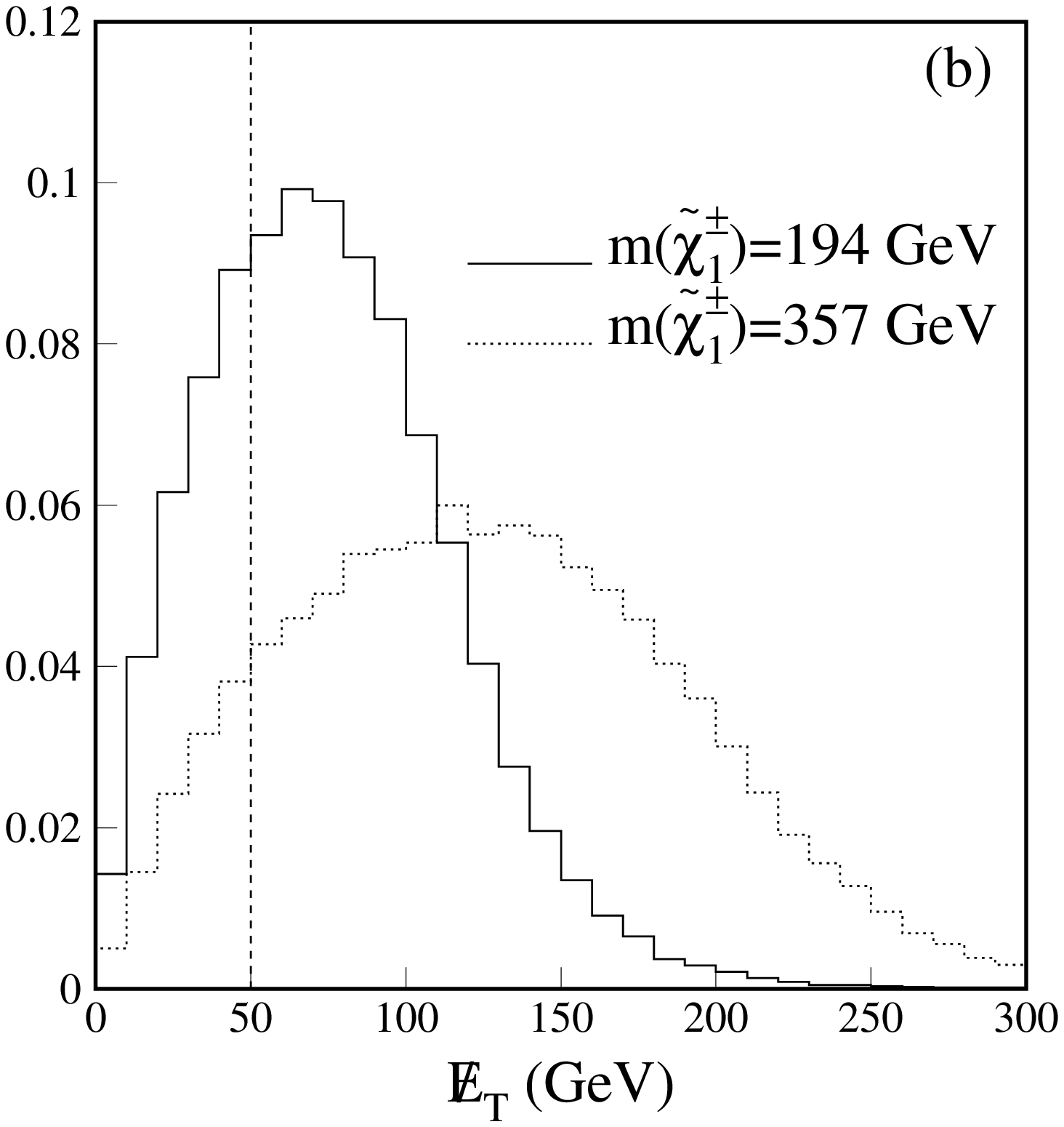}}
  \caption{Distributions of (a) photon $E_T$ and (b) event \protect\met\ for
           $\Lambda=80,\ 140$~TeV ($m_{\tilde\chi^\pm_1}=194,\ 357$~GeV) for 
           the models with a short-lived \protect\nlsp\ as the {\sc nlsp}. 
           The vertical dashed lines indicate the cutoffs. 
           All distributions are normalized to have unit area.}
  \label{fig:p1}
\end{figure}

\begin{figure}[htbp]
  \centerline{\epsfysize=3.5in\epsfbox{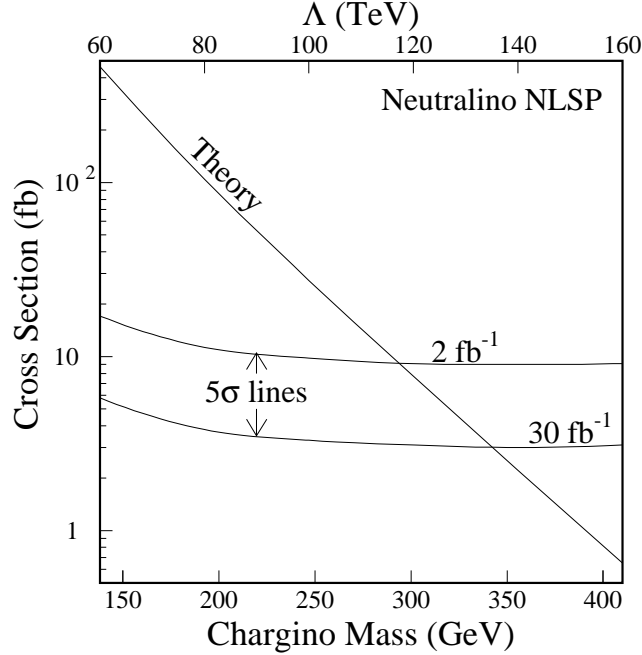}}
  \caption{The $5\sigma$ discovery cross section curves 
           as functions of mass of the lighter chargino (and the supersymmetry 
           breaking scale $\Lambda$) along with the theoretical cross sections 
           for the  model line 1. The {\sc nlsp} \protect\nlsp\ is assumed to 
           be short-lived. Two curves corresponding to integrated
           luminosities of 2 and 30~fb$^{-1}$ are shown.}
  \label{fig:p1lim}
\end{figure}

If the \nlsp\ decays in the vicinity of the production vertex, \ggmet\ events
are expected. The distributions of photon $E_T$ and event \met\ for 
$\Lambda=80,140$~TeV are shown in Fig.~\ref{fig:p1}. These events 
typically have high $E_T$ photons together with large transverse momentum 
imbalances, and therefore can be selected using the \ggmet\ criteria 
discussed in Section~\ref{sec:ggmet}. Table~\ref{tab:p1} shows the detection
efficiencies, significances along with the total theoretical cross 
sections of supersymmetry, chargino and neutralino masses for different 
values of $\Lambda$. Figure~\ref{fig:p1lim} compares the $5\sigma$ 
discovery cross sections $\sigma_{dis}$ with the theoretical cross sections 
expected from supersymmetry for two different values of \ldt\ as functions of 
the lighter chargino mass $m_{\tilde\chi^\pm_1}$ (and the supersymmetry 
breaking scale $\Lambda$). The lighter chargino with mass up to 290, 340~GeV 
can be discovered for \ldt=2, 30~fb$^{-1}$ respectively.

\begin{table}[htbp]
  \begin{tabular}{c|cccccc}
  $\Lambda$ (TeV)              &  60 &   80 &  100 &  120 &  140 &  160 \\ 
\hline
  $\sigma_{th}$ (fb)           & 464 &  105 &   27 &  7.7 &  2.2 &  0.7 \\ 
  $m_{\tilde\chi^\pm_1}$ (GeV) & 138 &  194 &  249 &  304 &  357 &  410 \\ 
  $m_{\tilde\chi^0_1}$ (GeV)   &  75 &  104 &  132 &  160 &  188 &  216 \\ 
\hline
  $\epsilon$ (\%)              & 16.1& 24.3 & 28.2 & 30.1 & 30.6 & 30.2 \\
  \protect\rsb\ (2 fb$^{-1}$)  & 136 &  46  &  14  &  4.2 &  1.2 &  0.4 \\ 
  \protect\rsb\ (30 fb$^{-1}$) & 400 & 137  &  41  &  12  &  3.6 &  1.1 \\
 \end{tabular}
  \caption{The supersymmetry cross section ($\sigma_{th}$), 
           $\tilde\chi^\pm_1$ and $\tilde\chi^0_1$ masses, detection 
           efficiency of the \protect\ggmet\ selection, and significances 
           for different values of $\Lambda$ for the models with a 
           short-lived \protect\nlsp\ as the {\sc nlsp} (model line 1). 
           The relative statistical error on the efficiency is typically
           2\%. The observable background cross section is 
           assumed to be 0.6~fb with a 20\% systematic uncertainty.}
  \label{tab:p1}
\end{table}

\begin{figure}[htbp]
  \centerline{\epsfysize=3.0in\epsfbox{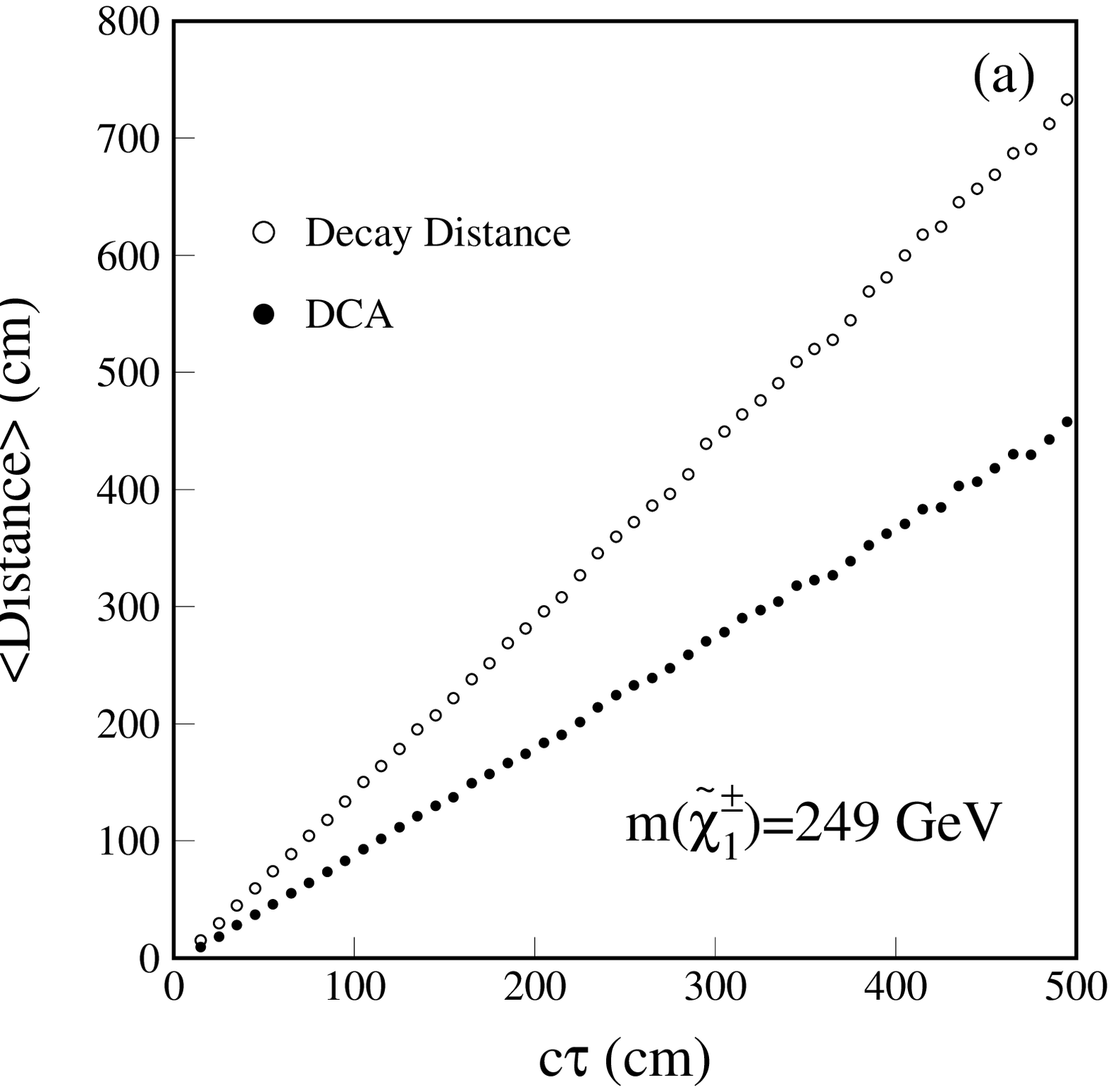}
              \epsfysize=3.0in\epsfbox{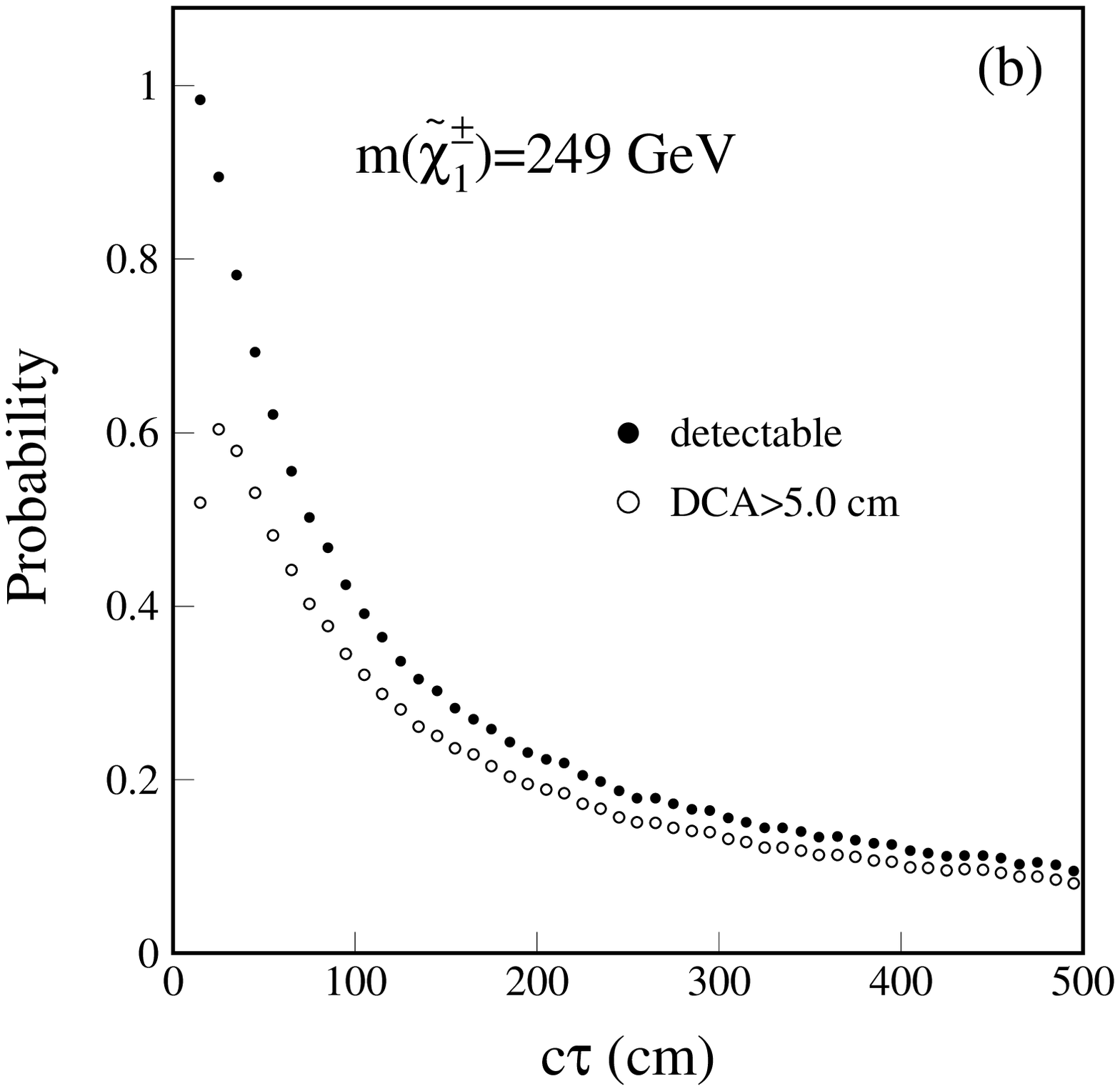}}
  \caption{(a) Average decay distance and {\sc dca} for different
           values of proper decay distance $c\tau$ of the {\sc nlsp} 
           \protect\nlsp. 
           (b) The probability that a photon decays within the tracking volume
           and the probability that such a photon has its 
           {\sc dca}$>5$~cm as functions of $c\tau$. Note that 
           $\Lambda=100$~TeV ($m_{\tilde\chi^\pm_1}=249$~GeV) for both plots.}
  \label{fig:p1dd}
\end{figure}

\subsubsection{Delayed $\tilde\chi^0_1\to\gamma\tilde G$ Decay}
\begin{figure}[htbp]
  \centerline{\epsfysize=3.0in\epsfbox{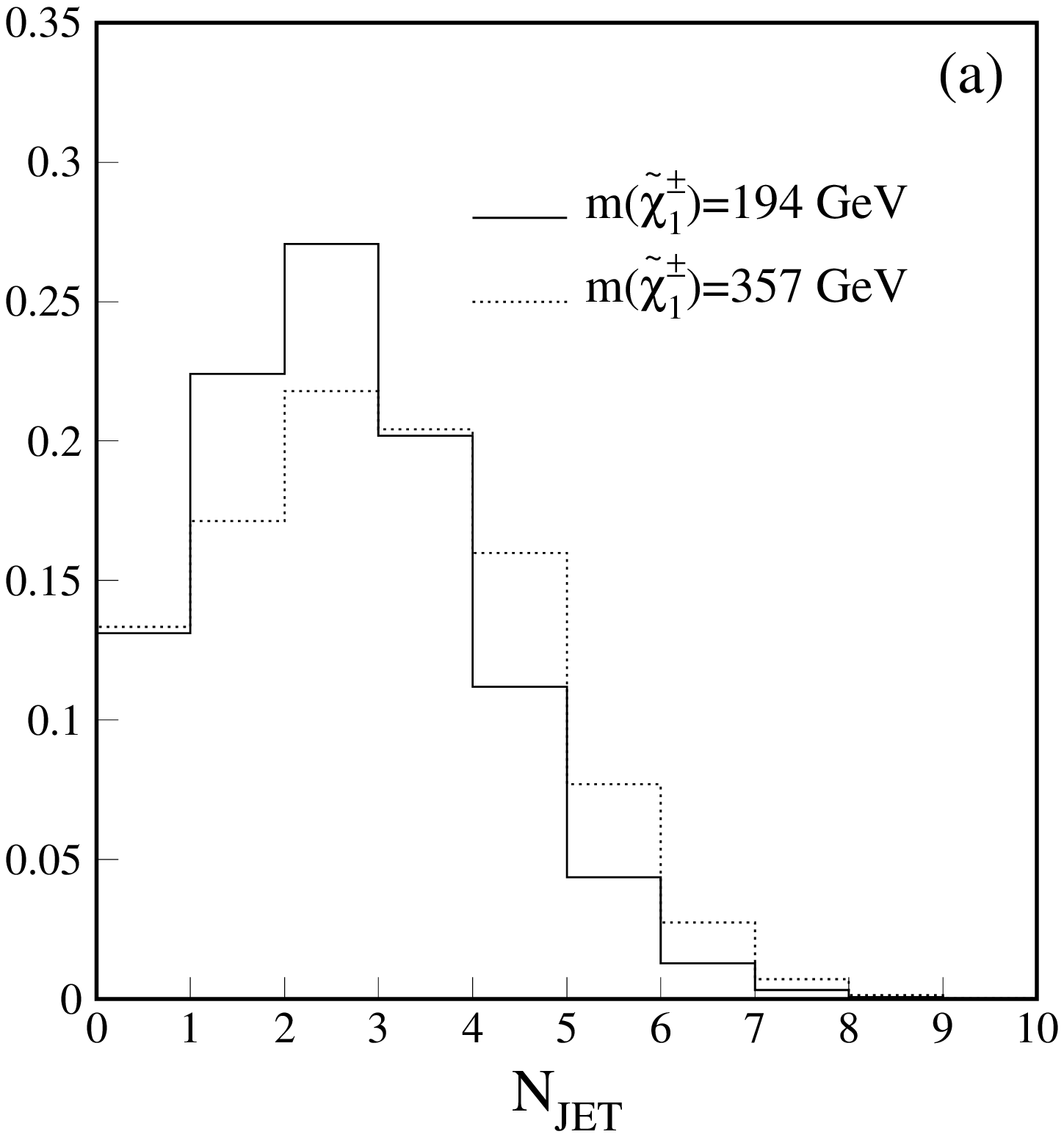}
              \epsfysize=3.0in\epsfbox{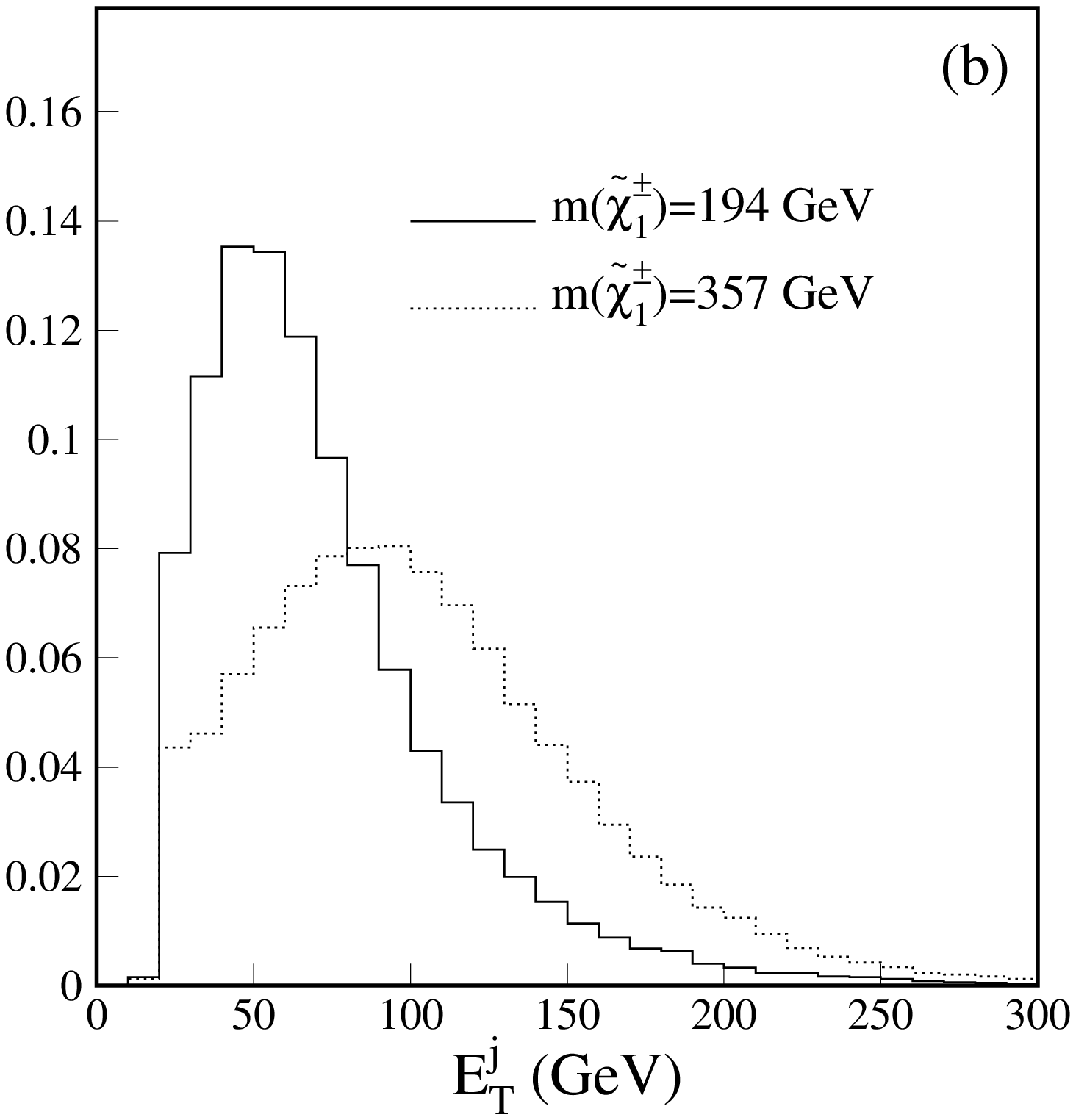}}
  \caption{Distributions of (a) jet multiplicity and (b) jet $E_T$
           for $\Lambda=80,\ 140$~TeV ($m_{\tilde\chi^\pm_1}=194,\ 357$~GeV) 
           for the models with \protect\nlsp\ as the {\sc nlsp}. All
           distributions are normalized to have unit area.} 
  \label{fig:p1jet}
\end{figure}

\begin{figure}[htbp]
  \centerline{\epsfysize=3.5in\epsfbox{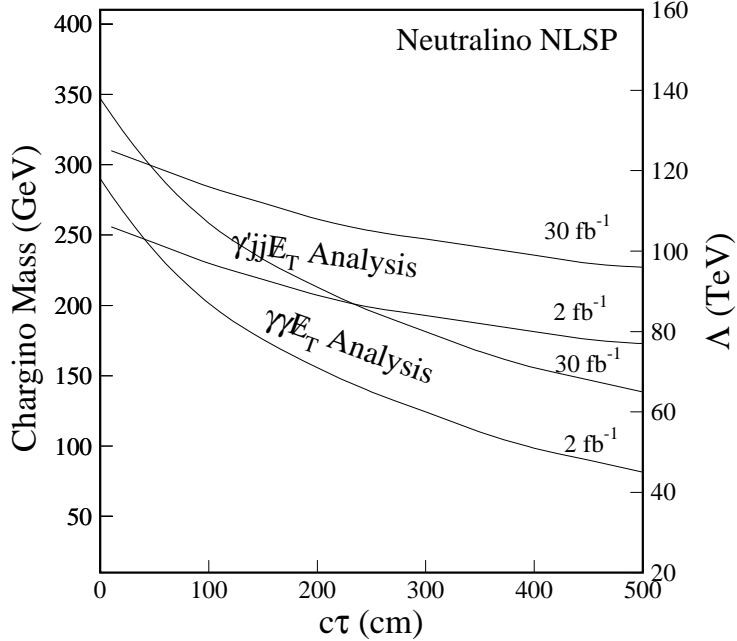}}
  \caption{The $5\sigma$ discovery reaches of the \protect\gdjjmet\ and
           the \ggmet\ analyses in $m_{\tilde\chi^\pm_1}$ and $\Lambda$
           as functions of the proper decay distance of the $\tilde\chi^0_1$ 
           {\sc nlsp} for \protect\ldt=2, 30~fb$^{-1}$.}
  \label{fig:p1dlim}
\end{figure}

If the \nlsp\ has a significant lifetime, the photon from its decay may not
point back to the primary vertex. If the decay occurs inside the tracking
volume of the D\O\ detector, the photon is expected to traverse standard 
electromagnetic detectors (the preshower detectors and the electromagnetic 
calorimeter). It, therefore, can be identified. 
However, if the decay occurs outside the tracking detector, the photon
identification is problematic. For this study, we assume that the photon
is identifiable if it is produced inside a cylinder defined by the
D\O\ tracking volume ($r<50$~cm and $|z|<120$~cm) and is lost if it is 
produced outside the cylinder. Figure~\ref{fig:p1dd}(a) shows the average decay
distance and distance of closest approach of the \nlsp\ as functions of its
proper decay length (c$\tau$) for $\Lambda=100$~TeV. Due to 
its heavy mass, the Lorentz boost for the \nlsp\ is typically small 
($\gamma\sim 1.5$). The probabilities that a photon is identifiable and 
that an identifiable photon has {\sc dca}$>5$~cm as functions of the 
$\tilde\chi^0_1$ proper decay distance 
$c\tau$ are shown in Fig.~\ref{fig:p1dd}(b), again for $\Lambda=100$~TeV.
Distributions for other $\Lambda$ values are similar. Figure~\ref{fig:p1jet}
shows jet multiplicity and $E_T$ distributions for $\Lambda=80,\
140$~TeV. Most of these events have large $E_T$ jets and thus can be 
selected using the \gdjjmet\ selection criteria discussed
above. The detection efficiencies and the expected significances of the 
\gdjjmet\ selection for $c\tau=50$~cm are tabulated in Table~\ref{tab:p1h} 
as an example. The estimated $5\sigma$ discovery reaches in $\Lambda$ and 
chargino mass for different values of $c\tau$ are shown in 
Fig.~\ref{fig:p1dlim} along with those expected from the \ggmet\ analysis.
As expected, the \ggmet\ analysis has a stronger dependence on $c\tau$ 
than the \gdjjmet\ analysis.

\begin{table}[htbp]
  \begin{tabular}{c|cccccc}
  $\Lambda$ (TeV)              &  60  &   80 &  100 &  120 &  140 &  160 \\ 
\hline
  $\sigma_{th}$ (fb)           & 464  &  105 &   27 &  7.7 &  2.2 &  0.7 \\ 
  $m_{\tilde\chi^\pm_1}$ (GeV) & 138  &  194 &  249 &  304 &  357 &  410 \\ 
  $m_{\tilde\chi^0_1}$ (GeV)   &  75  &  104 &  132 &  160 &  188 &  216 \\ 
\hline
  $\epsilon$ (\%)              & 11.2 & 23.6 & 31.1 & 33.3 & 33.2 &  32.1\\
  \protect\rsb\ (2 fb$^{-1}$)  &  93  &  44  & 15   & 4.6  & 1.3  & 0.4  \\ 
  \protect\rsb\ (30 fb$^{-1}$) & 278  & 133  & 45   &  14  & 3.9  & 1.2  \\
 \end{tabular}
  \caption{The detection efficiency of the \protect\gdjjmet\ selection, and
           the significances for different values of $\Lambda$ for the models 
           with \protect\nlsp\ as the {\sc nlsp}. The \protect\nlsp\ proper
           decay distance $c\tau$ is assumed to be 50~cm. The relative 
           statistical
           error on the efficiency is typically 2\%. The observable background
           cross section is assumed to be 0.6~fb with a 20\% systematic
           uncertainty.}
  \label{tab:p1h}
\end{table}

\subsection{Model Line 2: $\tilde\tau_1$ as the NLSP}
\begin{figure}[htbp]
  \centerline{\epsfysize=3.5in\epsfbox{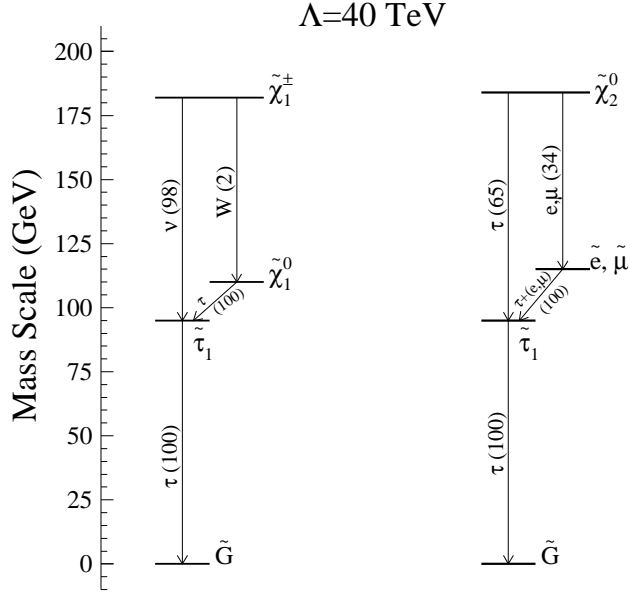}}
  \caption{Decay schematics of $\tilde\chi^\pm_1$ and $\tilde\chi^0_2$
           for $\Lambda=40$~TeV for the model line with a \protect\tlsp\ as
           the {\sc nlsp}. Percentage branching ratios for main decay modes
           are shown in parentheses.}
  \label{fig:p2br}
\end{figure}

If \tlsp\ (the lighter of the two mixed states of $\tilde\tau_R$ and
$\tilde\tau_L$) is lighter than \nlsp, all supersymmetric particles will 
cascade into the \tlsp\ which in turn will decay to $\tau\tilde G$ with 
a 100\% branching ratio. This class of models is defined by 
following parameter values:
$$N=2,\ \ \frac{M_m}{\Lambda}=3,\ \ \tan\beta=15,\ \ \mu>0$$
with varying $\Lambda$. Again, \cc\ and \cn\ dominate the production
cross section for $\Lambda\lsim 75$~TeV. For $\Lambda$ values
above 75~TeV, $\tilde\tau_1\tilde\tau_1$, $\tilde e_R\tilde e_R$, and
$\tilde\mu_R\tilde\mu_R$ productions become important. As an example,
branching ratios of $\tilde\chi^\pm_1$ and $\tilde\chi^0_2$ for
$\Lambda=40$~TeV are graphically displayed in Fig.~\ref{fig:p2br}.
In the following,
two cases corresponding to short-lived and quasi-stable $\tilde\tau_1$s
are discussed. It should be noted that the two analyses discussed below
are also sensitive to the case with a intermediate \tlsp\ lifetime.

\subsubsection{Prompt $\tilde\tau_1\to\tau\tilde G$ Decay}
\begin{figure}[htbp]
  \centerline{\epsfysize=3.0in\epsfbox{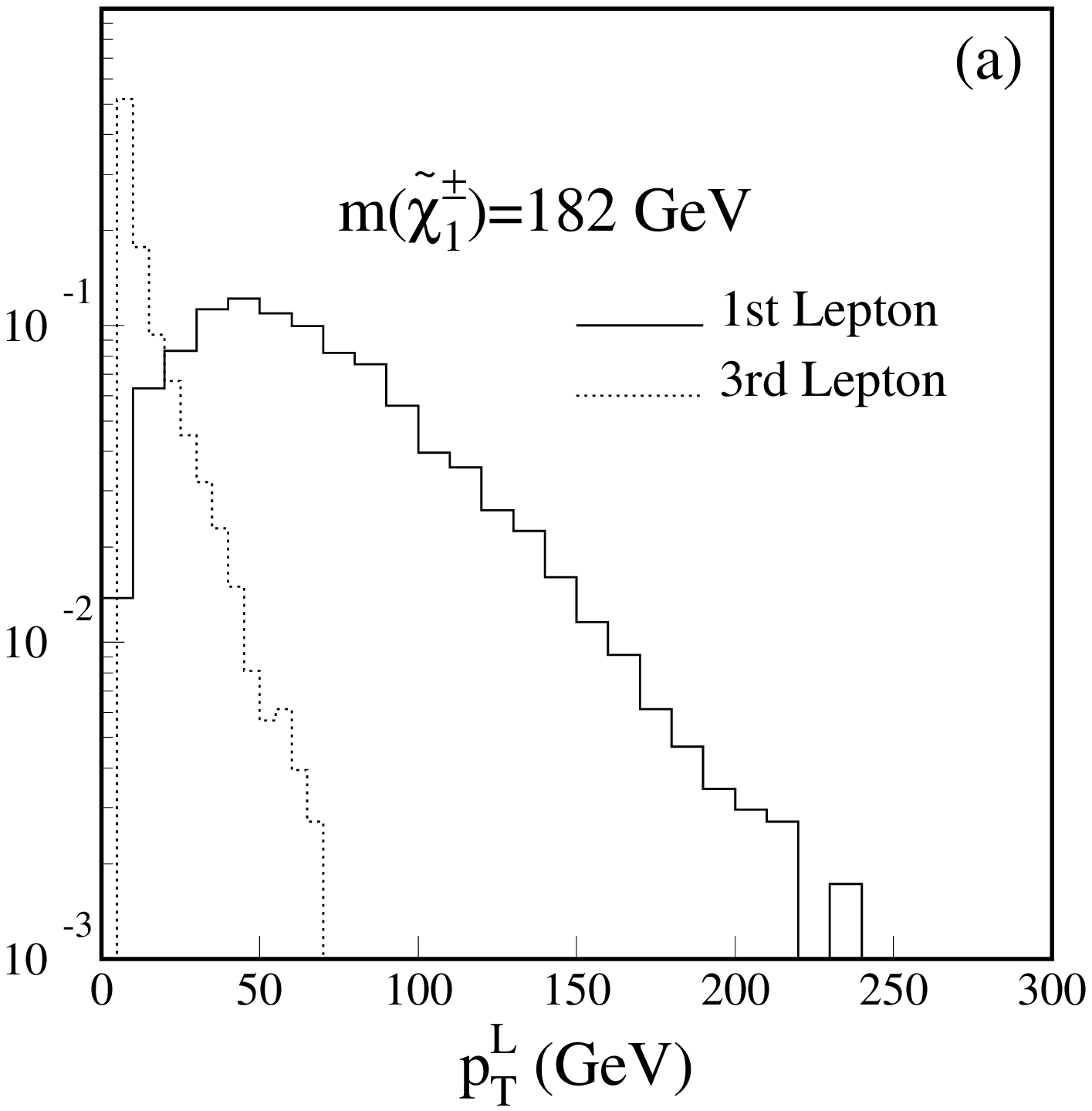}
              \epsfysize=3.0in\epsfbox{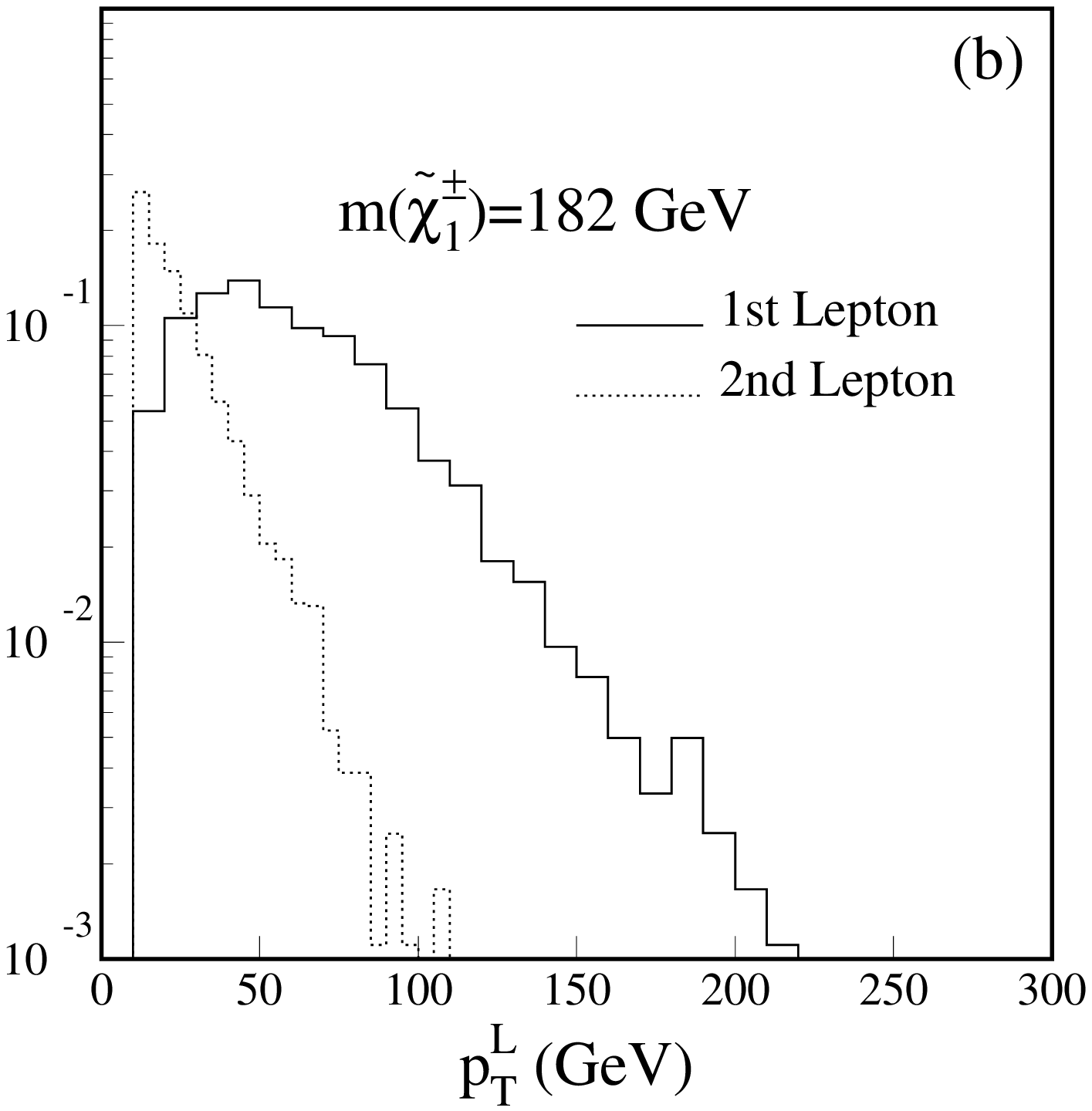}}
  \caption{Lepton $p_T$ distributions for (a) the \protect\lllj\ events and
           (b) the \protect\lljj\ events of the models with a short-lived 
           \protect\tlsp\ for $m_{\tilde\chi^\pm_1}=182$~GeV 
           ($\Lambda=40$~TeV). All distributions are normalized to unit area.}
  \label{fig:p2}
\end{figure}

\begin{figure}[htbp]
  \centerline{\epsfysize=3.5in\epsfbox{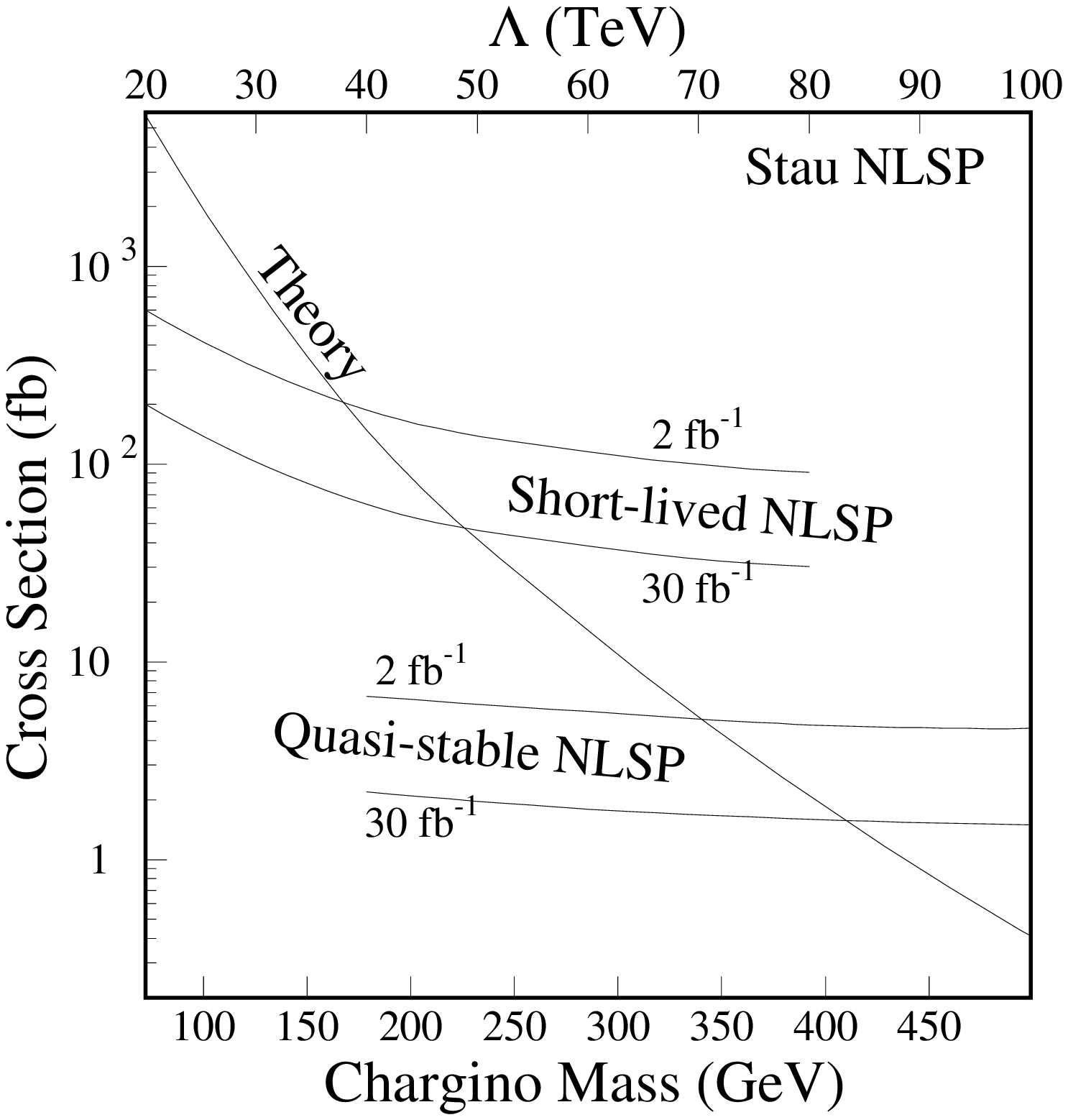}}
  \caption{The $5\sigma$ discovery cross section curves as functions of mass 
           of the lighter chargino and the supersymmetry breaking scale 
           $\Lambda$ for the model line 2 along with the theoretical 
           cross sections. The
           $5\sigma$ curves are shown for both short-lived and quasi-stable 
           {\sc nlsp}'s and for integrated luminosities of 2, 30~fb$^{-1}$. }
  \label{fig:p2lim}
\end{figure}

If the \tlsp\ is short-lived and decays in the vicinity of the production 
vertex ({\it i.e.} with a decay distance $\gamma c\tau\lsim 10$~cm),
anomalous $\tau$ production is expected from supersymmetry. Together with 
the $W^*/Z^*$ productions from the cascade decays of primary supersymmetric
particles, these events will give rise to \lllj\ and \lljj\ final states. The 
lepton $p_T$ distributions of the \lllj\ and \lljj\ events are shown
in Fig.~\ref{fig:p2}. Since most leptons are produced in $\tau$ decays, their 
$p_T$s are relatively soft. Table~\ref{tab:p2} shows the efficiencies of the 
\lllj\ and \lljj\ selection criteria for these events along with the 
theoretical 
cross sections, $\tilde\chi^\pm_1$ and $\tilde\tau_1$ masses. Note that the 
\lllj\ and \lljj\ criteria are orthogonal. The efficiencies are relatively
small largely due to the small branching ratio of the events to tri-leptons. 
We note that the total efficiencies shown in the table are somewhat
conservative. They do not take into account the migration of the \lllj\ 
events to the \lljj\ events due to inefficiency in the lepton 
identification. 
The $5\sigma$ discovery curves are shown in Fig.~\ref{fig:p2lim}. 
The lighter chargino with mass up to 160 and 230 GeV can be discovered for 
\ldt=2, 30~fb$^{-1}$.

The conventional wisdom is that this analysis should benefit from a $\tau$ 
identification. However, we doubt that it will have a dramatic impact on
the reach in the supersymmetry parameter space. Though a $\tau$ identification
could improve the efficiency for the signal, it will undoubtably come with
large backgrounds. Nevertheless, a $\tau$ identification is essential to 
narrow down 
theoretical models if an excess is observed in the tri-lepton final state.

\begin{table}[htbp]
  \begin{tabular}{c|cccc}
     $\Lambda$ (TeV)              &   20 &  40 &  60  &  80 \\ \hline
     $\sigma_{th}$ (fb)           & 5800 & 149 & 14.4 & 2.1 \\ 
     $m_{\tilde\chi^\pm_1}$ (GeV) &   72 & 182 & 289  & 394 \\
     $m_{\tilde\tau_1}$ (GeV)     &   54 &  99 & 147  & 196 \\ \hline
     \protect\lljj\ $\epsilon$ (\%) &  --  & 0.6 & 1.0  & 1.3 \\
     \protect\lllj\ $\epsilon$ (\%) &  0.5 & 1.0 & 1.6  & 2.0 \\
     Total  $\epsilon$ (\%)         &  0.5 & 1.6 & 2.6  & 3.3 \\ \hline
     \protect\rsb\ (2 fb$^{-1}$)    &  48  & 4.0 & 0.6  & 0.1 \\ 
     \protect\rsb\ (30 fb$^{-1}$)   & 140  &  12 & 1.8  & 0.3 \\ 
  \end{tabular}
  \caption{The theoretical cross section, $\tilde\chi^\pm_1$ and 
           $\tilde\tau_1$ masses, detection efficiency of \protect\lljj\
           and \protect\lllj\ selections, and significances 
           for different values of $\Lambda$ for the models with a short-lived
           $\tilde\tau_1$ as the {\sc nlsp}. The relative statistical 
           error on the efficiency is typically 25\%. The combined 
           \protect\lljj\ and 
           \protect\lllj\ background cross section is assumed 
           to be 0.7~fb with a 20\% systematic uncertainty.}
  \label{tab:p2}
\end{table}

\begin{figure}[htbp]
  \centerline{\epsfysize=3.0in\epsfbox{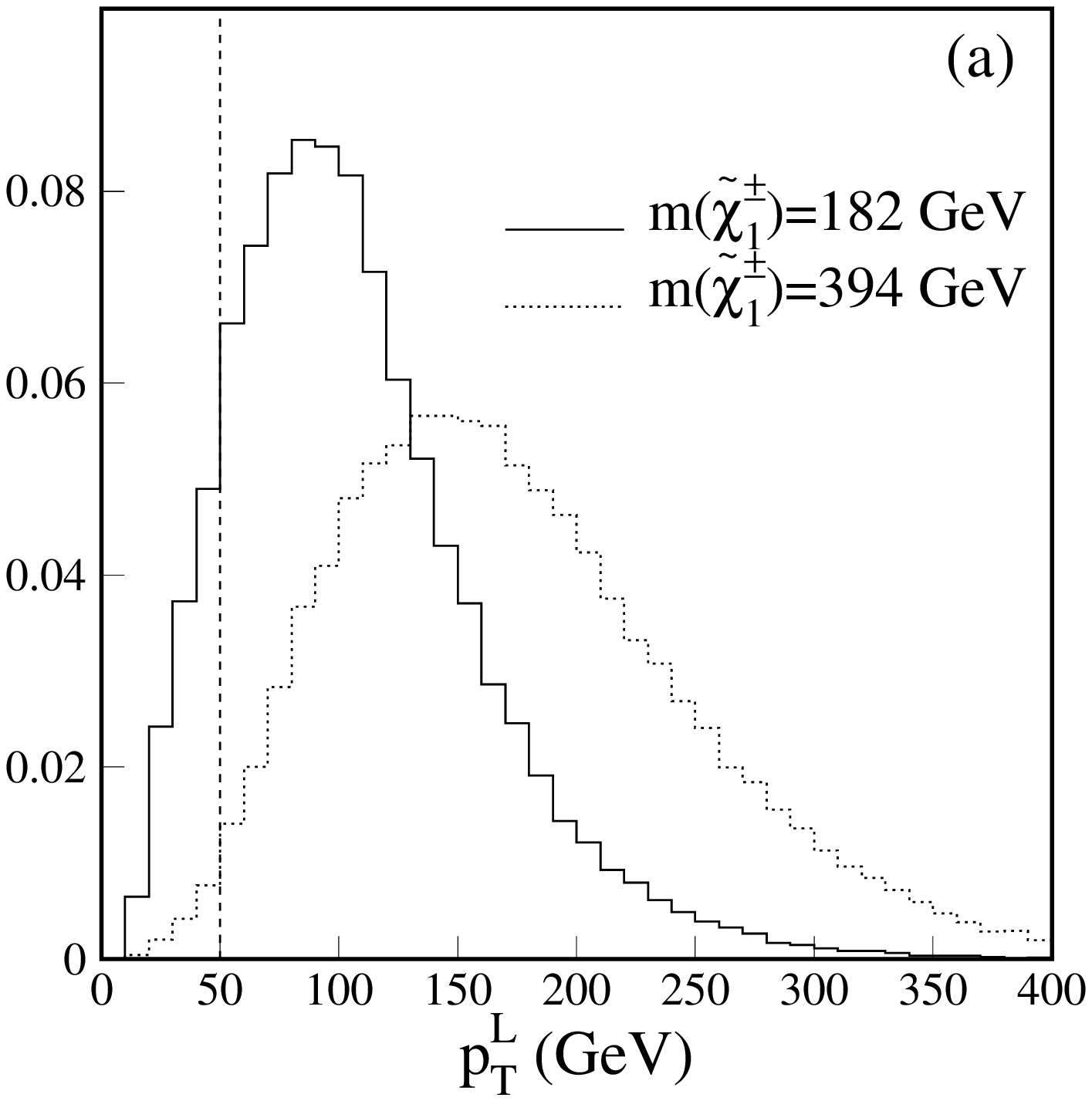}
              \epsfysize=3.0in\epsfbox{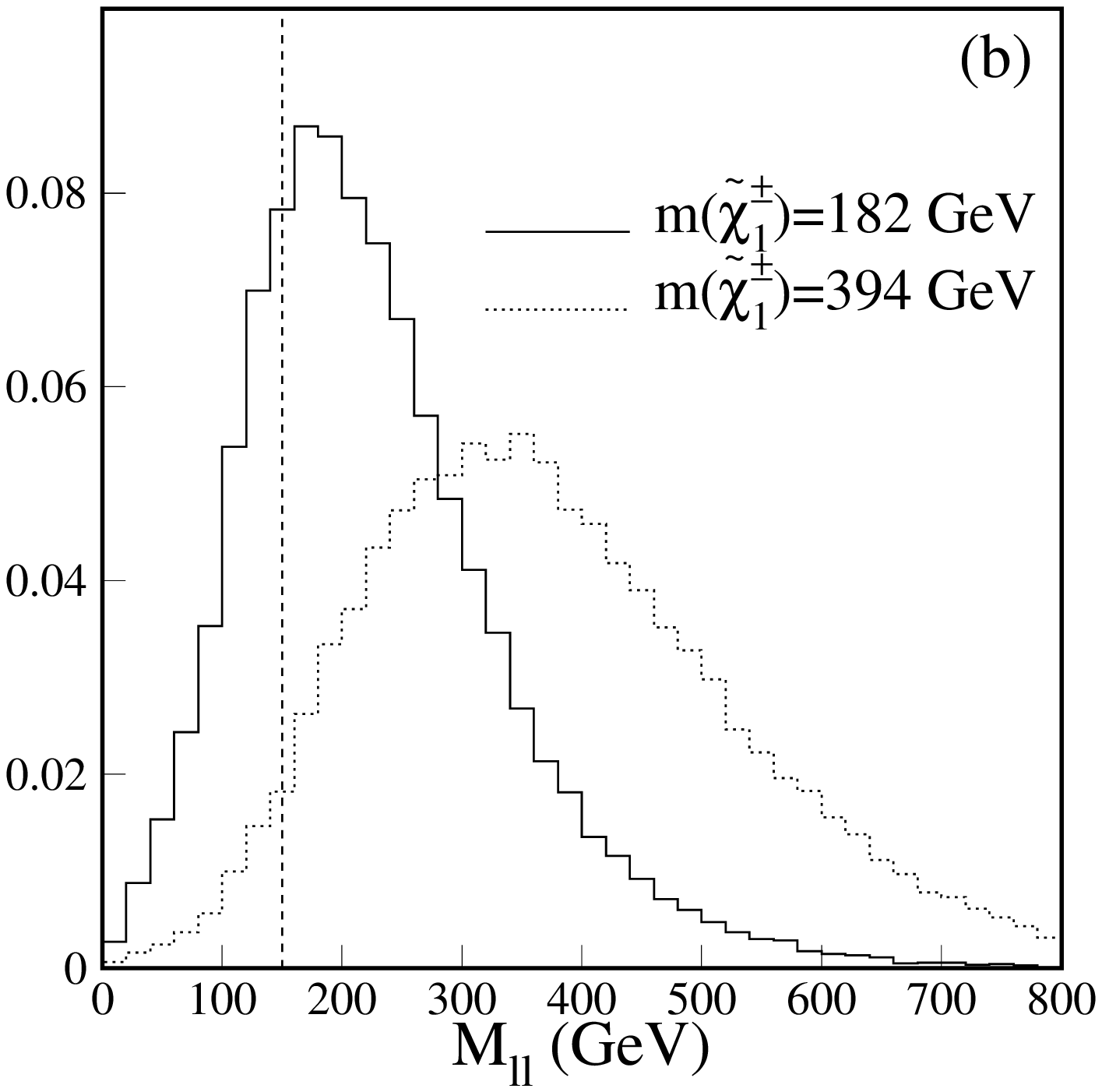}}
  \caption{The lepton $p_T$ (a) and di-lepton mass $M_{\ell\ell}$ (b) 
           distributions for the models with a quasi-stable \protect\tlsp\ 
           as the {\sc nlsp} for $\Lambda=40,\ 80$~TeV  
           ($m_{\tilde\chi^\pm_1}=182,\ 394$~GeV). The vertical dashed lines
           indicate the cutoffs. All distributions are normalized to 
           unit area.}
  \label{fig:p2h}
\end{figure}

\begin{figure}[htbp]
  \centerline{\epsfysize=3.2in\epsfbox{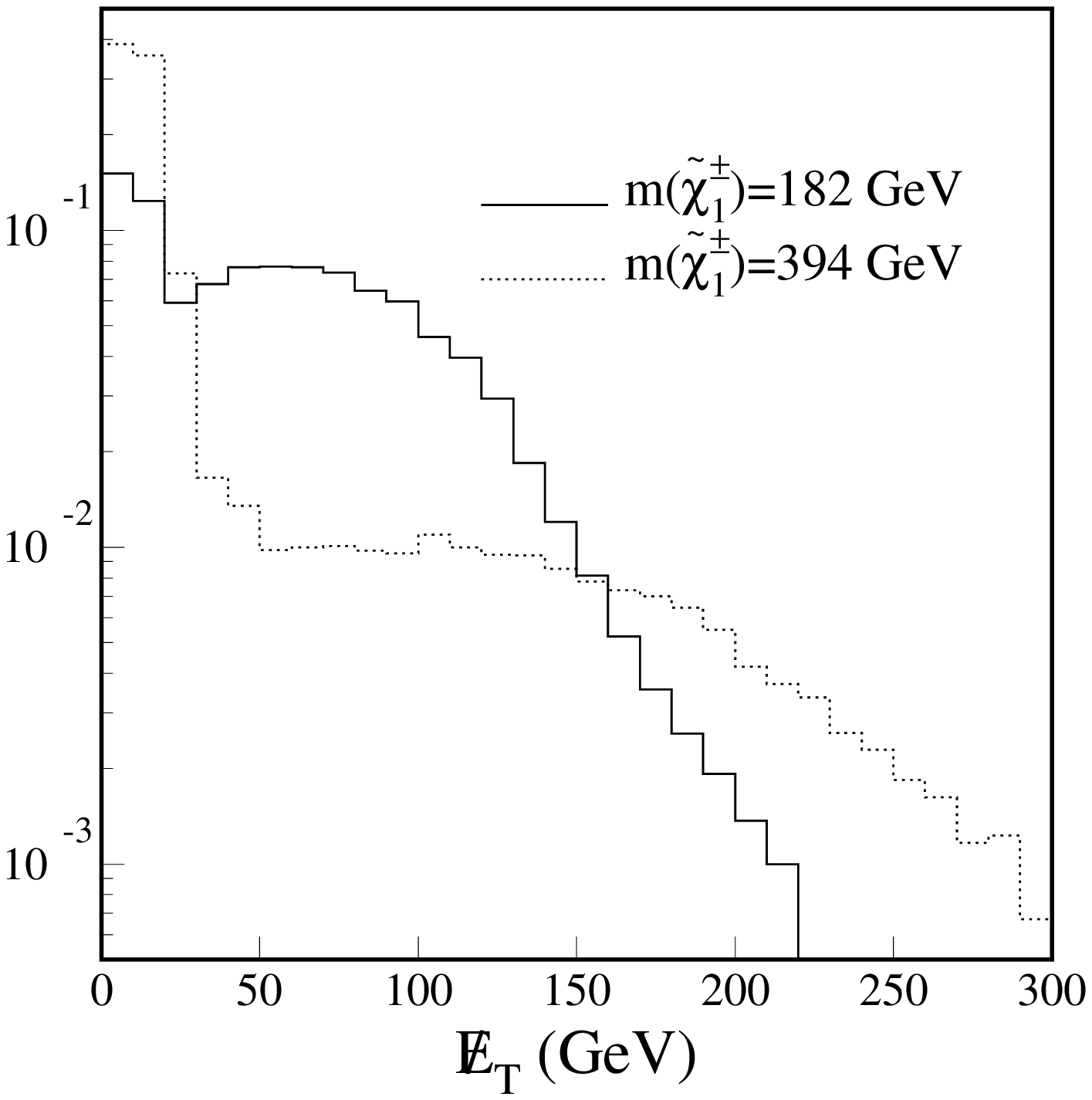}}
  \caption{The \protect\met\ distributions for the models with a quasi-stable
           \protect\tlsp\ as the {\sc nlsp} for $\Lambda=40,\ 80$~TeV
          ($m_{\tilde\chi^\pm_1}=182,\ 394$~GeV). The decays 
          $\tilde\chi^\pm_1\to\tilde\chi^0_1 W^\pm\to\tilde\tau_1\tau W^\pm$ 
          and $\tilde\chi^0_2\to e\tilde e,\ \mu\tilde\mu$ contribute to events 
          with small \protect\met\ while the decays
          $\tilde\chi^\pm_1\to\tilde\tau_1\nu$ and 
          $\tilde\chi^0_2\to\tau\tilde\tau_1$ are the source
          for events with large \protect\met.
          }
  \label{fig:p2hmet}
\end{figure}

\subsubsection{Quasi-stable $\tilde\tau_1$}
If the \tlsp\ has a long lifetime (quasi-stable) and decays outside the 
detector ($\gamma c\tau$ greater than $\sim 3$~m), 
it will appear in the detector like a muon with the exception of
a large ionization energy loss. The signature 
is, therefore, two high $p_T$ `muons' with large $dE/dx$ values.
These events can be selected using the criteria described in 
Section~\ref{sec:lldedx}. The expected $p_T$ distributions
of the \tlsp\ for two different values of $\Lambda$ are shown in 
Fig.~\ref{fig:p2h}(a). The cut of $p_T>50$~GeV of the \lldedx\ selection 
is efficient for the signal while it is expected to reduce backgrounds 
significantly. The typical invariant mass of the two `muons' (assuming 
massless) is very large as shown in Fig.~\ref{fig:p2h}(b).
A $M_{\ell\ell}>150$~GeV requirement does little harm to the signals.
Due to its large mass, the \tlsp\ is expected to move slowly. However since
most of the \tlsp's are produced in the decays of massive $\tilde\chi^\pm_1$s
and $\tilde\chi^0_2$s, the average speed $\beta(\equiv v/c)$ is relatively 
large. It is around 0.7 for the $\Lambda$ values studied.
Note that the $\beta$ distribution is very similar to that shown in
Fig.~\ref{fig:p3h}(b) for the models with $\tilde\ell$ as the Co-{\sc nlsp}.
Nevertheless, the not-so-slow moving \tlsp's are expected to deposit large 
ionization energies in the detector, differentiating them from other high 
$p_T$ MIP particles. Since the backgrounds for the requirements $p_T>50$~GeV 
and $M_{\ell\ell}>150$~GeV are already small, it pays to have a $dE/dx$ 
requirement with a relatively high efficiency for the signal and a 
resonable rejection for the MIP particles. 
The \protect\met\ distribution of these events 
as shown in Fig.~\ref{fig:p2hmet} shows two distinct regions: small and
large \met. The decays $\tilde\chi^\pm_1\to\tilde\chi^0_1 W\to 
\tilde\tau_1\tau W$ and $\tilde\chi^0_2\to e\tilde e,\ \mu\tilde\mu$ 
contribute to events with small \met.  The decays 
$\tilde\chi^\pm_1\to \tilde\tau_1\nu$ and 
$\tilde\chi^0_2\to\tau\tilde\tau_1$ are responsible for events with large \met. 
The detection efficiencies and the expected significances of the 
\lldedx\ selection for different values of $\Lambda$ are tabulated 
in Table~\ref{tab:p2h}. The high efficiency is largely due to the high
momentum expected for the quasi-stable $\tilde\tau_1$.
The $5\sigma$ discovery curves are shown in Fig.~\ref{fig:p2lim} for two 
values of \ldt. The lighter chargino with mass up to 
340, 410~GeV and the \tlsp\ with mass up to 160, 200~GeV can be discovered 
for the two integrated luminosities respectively.

\begin{table}[htbp]
  \begin{tabular}{c|cccc}
     $\Lambda$ (TeV)              &   40 &  60  &   80 &  100 \\ \hline
     $\sigma_{th}$ (fb)           &  149 & 14.4 &  2.1 &  0.4 \\ 
     $m_{\tilde\chi^\pm_1}$ (GeV) &  182 & 289  &  394 &  499 \\
     $m_{\tilde\tau_1}$ (GeV)     &   99 & 147  &  196 &  246 \\ \hline
     $\epsilon$ (\%)              & 37.4 & 44.6 & 51.6 & 54.9 \\
     \protect\rsb\ (2 fb$^{-1}$)  &  112 & 12   &  2.1 &  0.5 \\ 
     \protect\rsb\ (30 fb$^{-1}$) &  341 & 40   &  6.7 &  1.4 \\ 
  \end{tabular}
  \caption{The theoretical cross section, $\tilde\chi^\pm_1$ and 
           $\tilde\tau_1$ masses, detection efficiency of the \protect\lldedx\
           selection, and significances for 
           different values of $\Lambda$ for the models with a quasi-stable
           $\tilde\tau_1$ {\sc nlsp}. The relative statistical error on the
           efficiency is typically 1\%. The background cross section 
           is assumed to be 0.5~fb with an uncertainty of 20\%.}
  \label{tab:p2h}
\end{table}

\subsection{Model Line 3: $\tilde\ell$ as the Co-NLSP}
\begin{figure}[htbp]
  \centerline{\epsfysize=3.5in\epsfbox{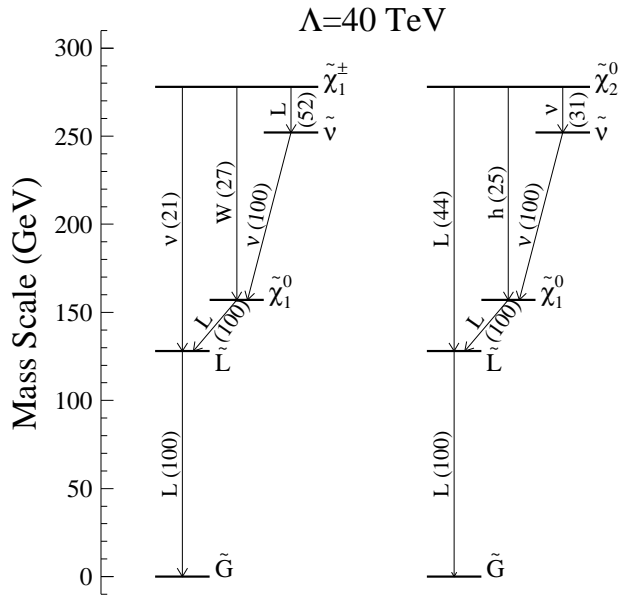}}
  \caption{Decay schematics of $\tilde\chi^\pm_1$ and $\tilde\chi^0_2$
           for $\Lambda=40$~TeV for the model line with scalar leptons
           as the Co-{\sc nlsp}. Percentage branching ratios for main
           decay modes are shown in parentheses.}
  \label{fig:p3br}
\end{figure}

For some regions of the parameter space, three light sleptons 
($\tilde e_R$, $\tilde\mu_R$, and $\tilde\tau_1$) are essentially degenerate
in mass and they can be lighter than \nlsp. As a result, the 
sleptons ($\tilde\ell\equiv \tilde\tau_1,\tilde e_R,\tilde\mu_R$) effectively 
share the role of the {\sc nlsp}. 
Quantitative studies of this type of models are done for following
GMSB parameter values:
$$N=3,\ \ \frac{M_m}{\Lambda}=3,\ \ \tan\beta=3,\ \ \mu>0$$
with again varying $\Lambda$ values. For small values of $\Lambda$, \cc\ 
and \cn\ dominate the production cross section. As shown in 
Fig.~\ref{fig:p3br}, \cc\ and \cn\ production will yield events with
multileptons in the final state.
The slepton pair production surpasses chargino-neutralino production 
if $\Lambda\gsim 50$~TeV. The total
supersymmetry cross section for several different $\Lambda$ values can be
found in Table~\ref{tab:p3}. The lifetime of $\tilde\ell$ determines the
event topology. In the following, we discuss the cases with short-lived 
and quasi-stable $\tilde\ell$s. Again, the analyses should also be sensitive
to the $\tilde\ell$ {\sc nlsp} with a intermediate lifetime.

\subsubsection{Prompt $\tilde\ell\to\ell\tilde G$ Decay}
\begin{figure}[htbp]
  \centerline{\epsfysize=3.0in\epsfbox{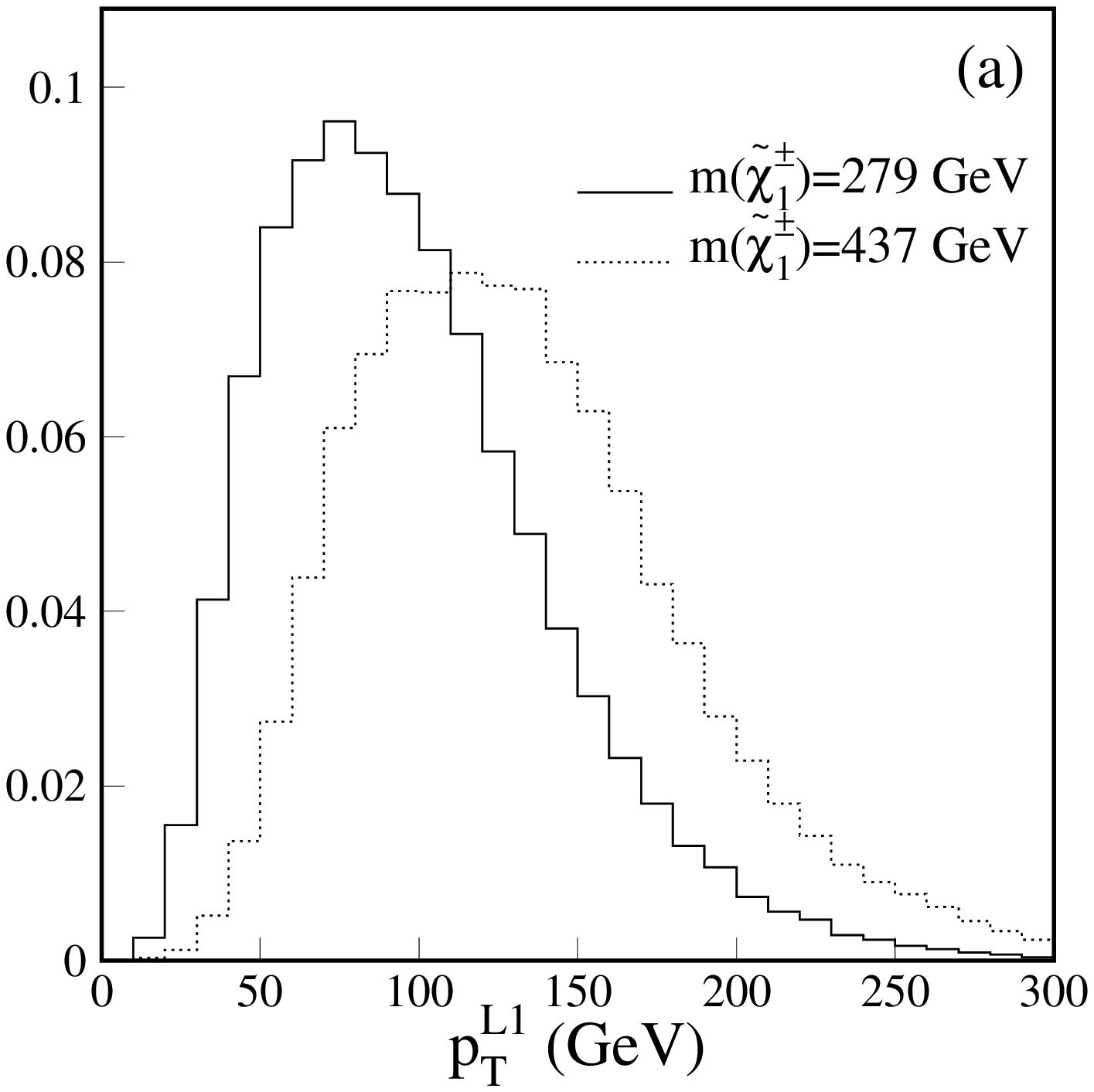}
              \epsfysize=3.0in\epsfbox{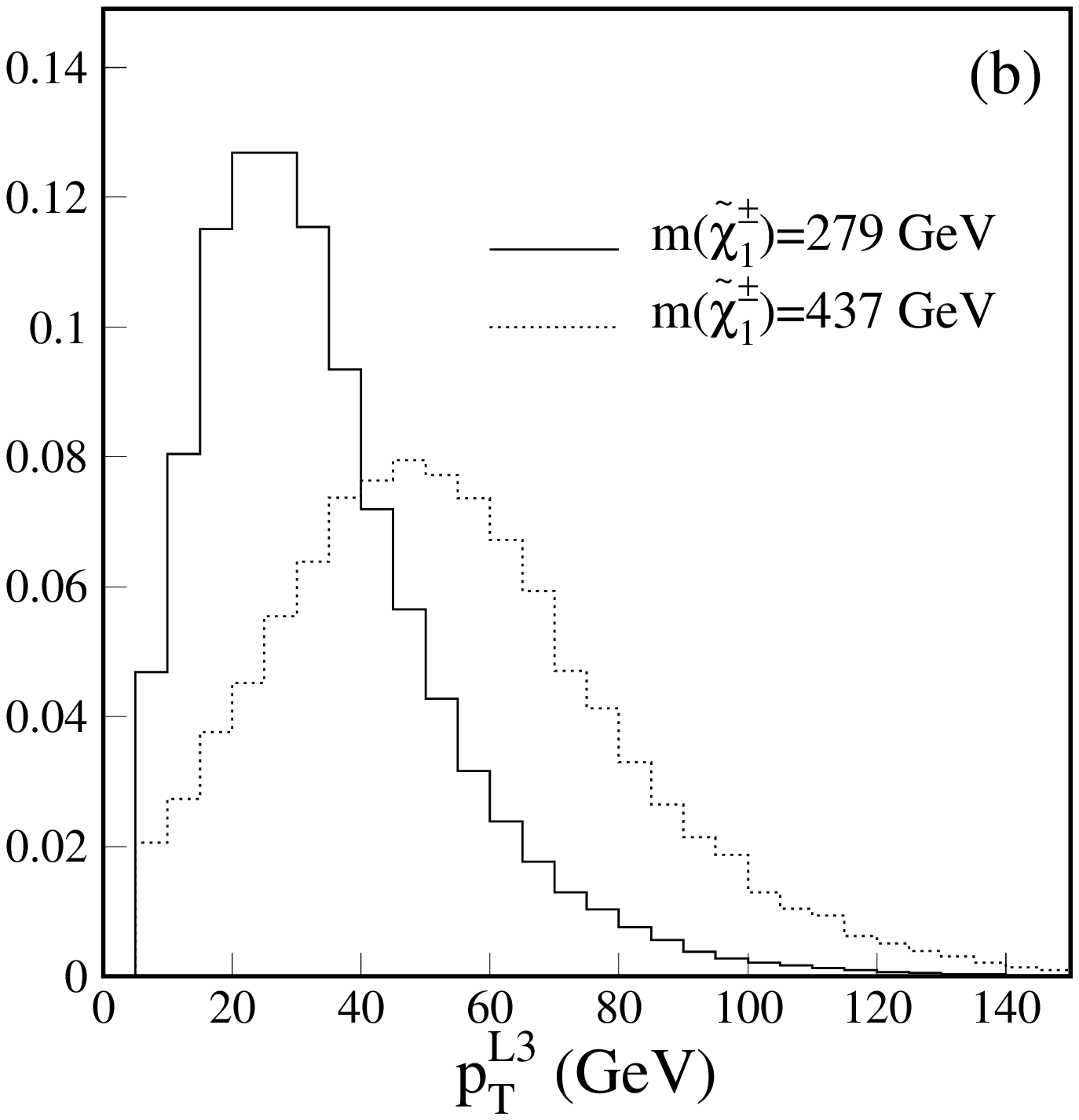}}
  \caption{The $p_T$ distributions of (a) the leading lepton and (b) the third
           lepton for the models with a short-lived $\tilde\ell$ as 
           the Co-{\sc nlsp} for $m_{\tilde\chi^\pm_1}=279,\ 437$~GeV 
           ($\Lambda=40,\ 60$~TeV). Note that the $p_T$ requirement is 
           15~GeV for the leading lepton and 5~GeV for the non-leading leptons.
           All distributions are normalized to unit area.}
  \label{fig:p3}
\end{figure}

\begin{figure}[htbp]
  \centerline{\epsfysize=3.5in\epsfbox{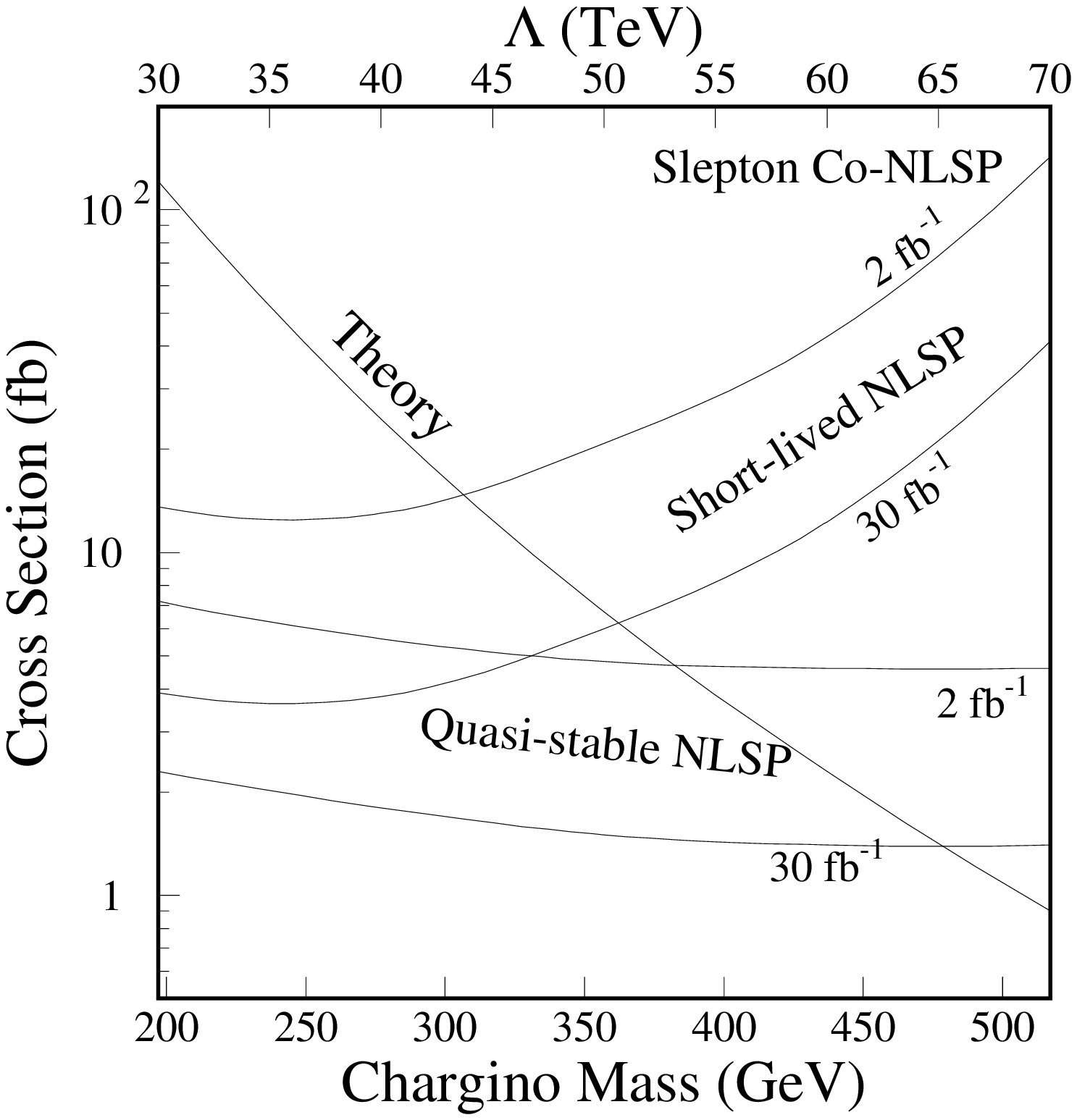}}
  \caption{The $5\sigma$ discovery cross section curves as functions of 
           mass of the lighter chargino and the supersymmetry breaking scale 
           $\Lambda$ for the model line 3 along with the theoretical cross 
           sections. The $5\sigma$ curves are shown for both short-lived and 
           quasi-stable $\tilde\ell$ Co-{\sc nlsp}'s and for integrated 
           luminosities of 2, 30~fb$^{-1}$. }
  \label{fig:p3lim}
\end{figure}

If the decay $\tilde\ell\to\ell\tilde G$ is prompt ($\gamma c\tau\lsim 10$~cm),
 \llmet\ events are
expected from supersymmetry. Unfortunately, this final state is swamped by
backgrounds from the Standard Model processes such as $t\bar{t}$, $WW$, 
$WZ$ and $ZZ$ productions as well as from $W+{\rm jets}$ production with
one of the jets misidentified as a lepton. However we note that 
these events typically have multiple leptons in the final state and most of
them are in the central pseudorapidity region with good lepton identification.
Apart from those from $\tilde\ell$ decays, leptons are also expected from 
$W^*$'s and $Z^*$'s produced in the cascade decays of $\tilde\chi^\pm_1$ and 
$\tilde\chi^0_2$ of supersymmetry originated events. Therefore, they can be 
selected using the \lllj\ criteria. The $p_T$ distributions
of the leading lepton and the third lepton of these events are shown in 
Fig.~\ref{fig:p3}(a). Since most of the leading leptons are produced in 
the direct decays of heavy $\tilde\ell$'s, its $p_T$ spectrum is relatively 
hard as shown in the figure. The detection efficiencies and the expected 
significances are summarized in Table~\ref{tab:p3}. The reduction in the 
relative cross section of the tri-lepton producing \cc\ and \cn\ processes 
is responsible for the decrease in efficiency as $\Lambda$ increases. 
For $\Lambda\gsim 50$~TeV, the $\tilde\ell\tilde\ell$ production cross 
section surpasses that of the \cc\ and \cn. With the 
$\tilde\ell\to\ell\tilde G$
decay, $\tilde\ell\tilde\ell$ events will result in a high $p_T$ 
$\ell\ell$\met\ final state. We note that the 
improvement by adding the \lljj\ selection is minimal in this case. 
The $5\sigma$ discovery curves are compared with the theoretical cross 
sections in Fig.~\ref{fig:p3lim}. With integrated luminosities of 2, and 
30~fb$^{-1}$, the lighter chargino with mass up to 310 and 360 GeV can be 
discovered respectively.

\begin{table}[htbp]
  \begin{tabular}{c|ccccc}
   $\Lambda$ (TeV)                &  30  &   40 &  50  &  60  &  70  \\ \hline
   $\sigma_{th}$ (fb)             & 121  & 24.5 & 6.7  & 2.3  & 0.9  \\ 
   $m_{\tilde\chi^\pm_1}$ (GeV)   & 197  &  279 & 358  & 437  & 517  \\
   $m_{\tilde\ell}$ (GeV)         &  99  &  128 & 158  & 188  & 218  \\ \hline
                  $\epsilon$ (\%) & 14.7 & 15.4 &  9.6 & 4.7  &  1.4 \\
   \protect\rsb\ (2 fb$^{-1}$)    &  44  & 9.4  &  1.6 & 0.3 &  -- \\ 
   \protect\rsb\ (30 fb$^{-1}$)   & 152  & 32   &  5.5 & 0.9 &  0.1 \\ 
  \end{tabular}
  \caption{The theoretical cross section, $\tilde\chi^\pm_1$ and $\tilde\ell$ 
           masses, detection efficiency of the \protect\lllj\  
           selection criteria, and significances for different values of 
           $\Lambda$ for the models with short-lived $\tilde\ell$'s as the 
           Co-{\sc nlsp}. The efficiencies typically have a relative 
           statistical uncertainty of 4\%. The observable background cross
           section is assumed to be 0.3~fb with a 20\% systematic uncertainty.
           }
  \label{tab:p3}
\end{table}

\begin{figure}[htbp]
  \centerline{\epsfysize=3.0in\epsfbox{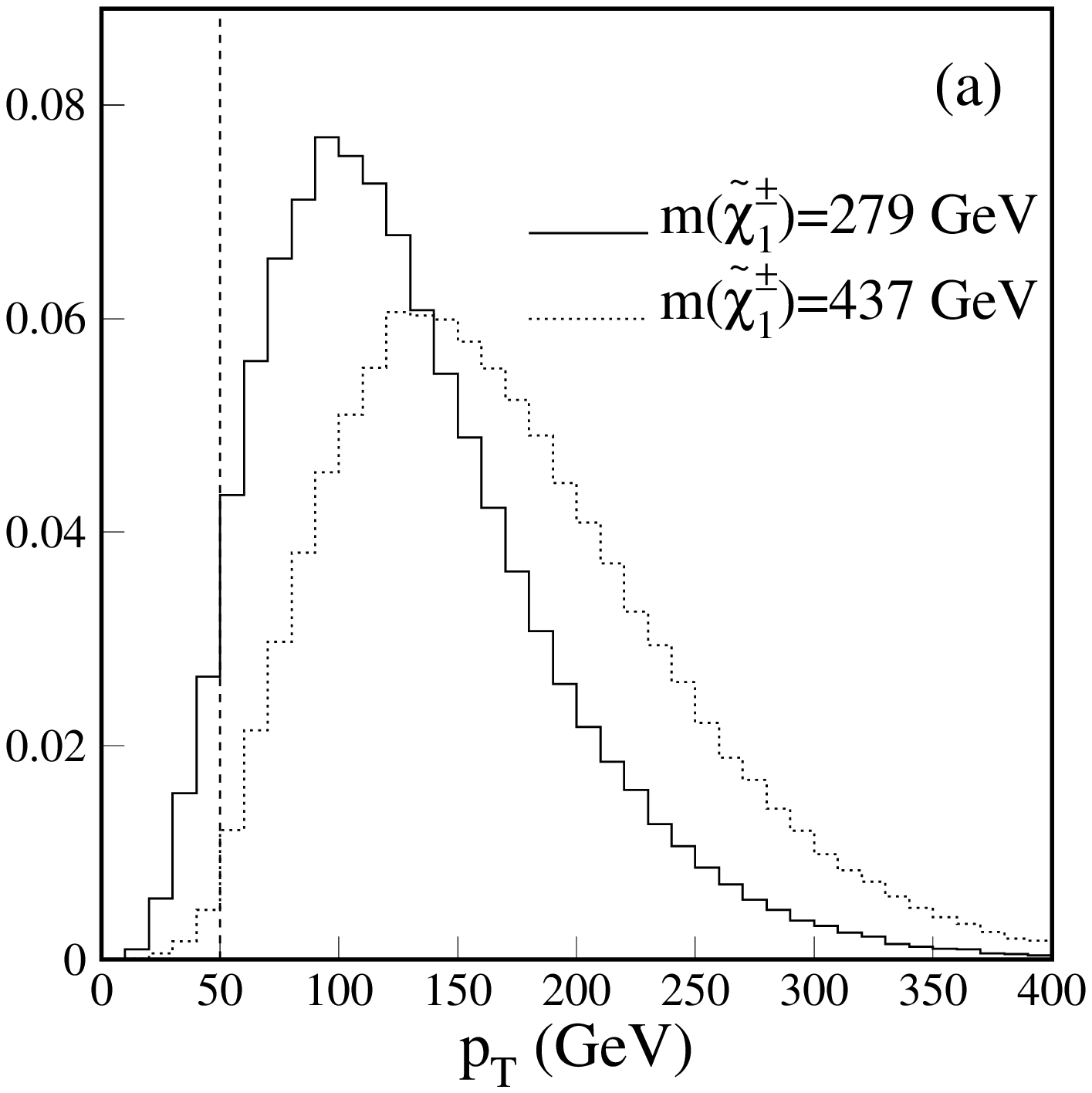}
              \epsfysize=3.0in\epsfbox{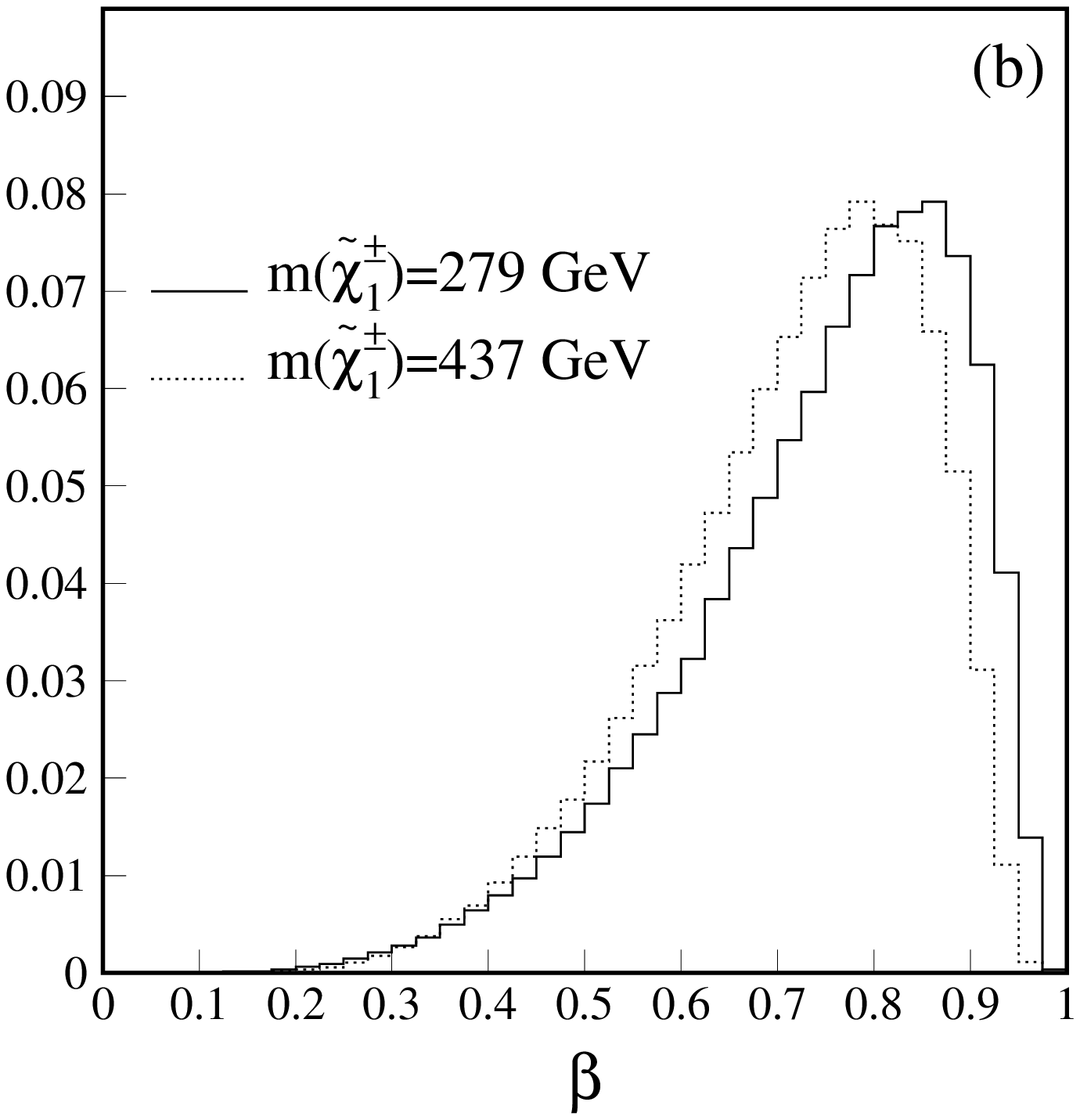}}
  \caption{The lepton $p_T$ (a) and the {\sc nlsp} speed $\beta$ (b) 
           distributions expected for models with a quasi-stable 
           $\tilde\ell$ as the Co-{\sc nlsp} for 
           $m_{\tilde\chi^\pm_1}=279,\ 437$~GeV ($\Lambda=40, 60$~TeV), 
           where $\beta$ is measured in the unit of the speed of
           light $c$. The vertical dashed line indicates the cutoff. 
           All distributions are normalized to unit area.}
  \label{fig:p3h}
\end{figure}

\subsubsection{Quasi-stable $\tilde\ell$}
If the $\tilde\ell$ has a long lifetime, it can decay outside the detector
($\gamma c\tau\gsim 3$~m).
In this case, the $\tilde\ell$ will appear in the detector like a `muon' 
except that the ionization energy loss will be large. This signature is 
identical
to that of a quasi-stable $\tilde\tau_1$ discussed above. Therefore, the
signal events can be identified using the \lldedx\ selection. 
The expected $p_T$ and $\beta$ distributions of the $\tilde\ell$ for
$\Lambda=40,\ 60$~TeV are shown in Fig.~\ref{fig:p3h}. The $\tilde\ell$s 
typically have very large $p_T$ and are mostly central. For example, about
90\% of the $\tilde\ell$s are in central pseudorapidity region with the
tracking coverage for the case of $\Lambda=70$~TeV. Table~\ref{tab:p3h} 
shows the detection efficiencies and the expected significances for different 
$\Lambda$ values. The $5\sigma$ discovery curves are shown in 
Fig.~\ref{fig:p3lim}. The lighter chargino mass discovery reach is
about 390 GeV for \ldt=2~fb$^{-1}$ and 480 GeV for \ldt=30~fb$^{-1}$.

\begin{table}[htbp]
  \begin{tabular}{c|ccccc}
     $\Lambda$ (TeV)              &  30  &   40 &   50 &  60  &   70 \\ \hline
     $\sigma_{th}$ (fb)           & 121  & 24.5 &  6.7 & 2.3  &  0.9 \\ 
     $m_{\tilde\chi^\pm_1}$ (GeV) & 197  &  279 &  358 & 437  &  517 \\
     $m_{\tilde\ell}$ (GeV)       &  99  &  128 &  158 & 188  &  218 \\ \hline
     $\epsilon$ (\%)              & 34.8 & 45.2 & 52.6 & 54.9 & 55.1 \\
     \protect\rsb\ (2 fb$^{-1}$)  &  83  &  22  &  7.0 &  2.6 &  1.0 \\ 
     \protect\rsb\ (30 fb$^{-1}$) & 257  &  68  &  21  &  7.8 &  3.1 \\ 
  \end{tabular}
  \caption{The theoretical cross section, $\tilde\chi^\pm_1$ and $\tilde\ell$ 
           masses, detection efficiency of the \protect\lldedx\ selection, 
           and significances for different values of $\Lambda$ for the 
           models with quasi-stable $\tilde\ell$'s as the Co-{\sc nlsp}.
           The relative statistical error on the efficiency is typically 1\%.
           The background cross section is assumed to be 0.5~fb with 
           a systematic uncertainty of 20\%.}
  \label{tab:p3h}
\end{table}

\subsection{Model Line 4: $\tilde h$ as the NLSP}
\begin{figure}[htbp]
  \centerline{\epsfysize=3.0in\epsfbox{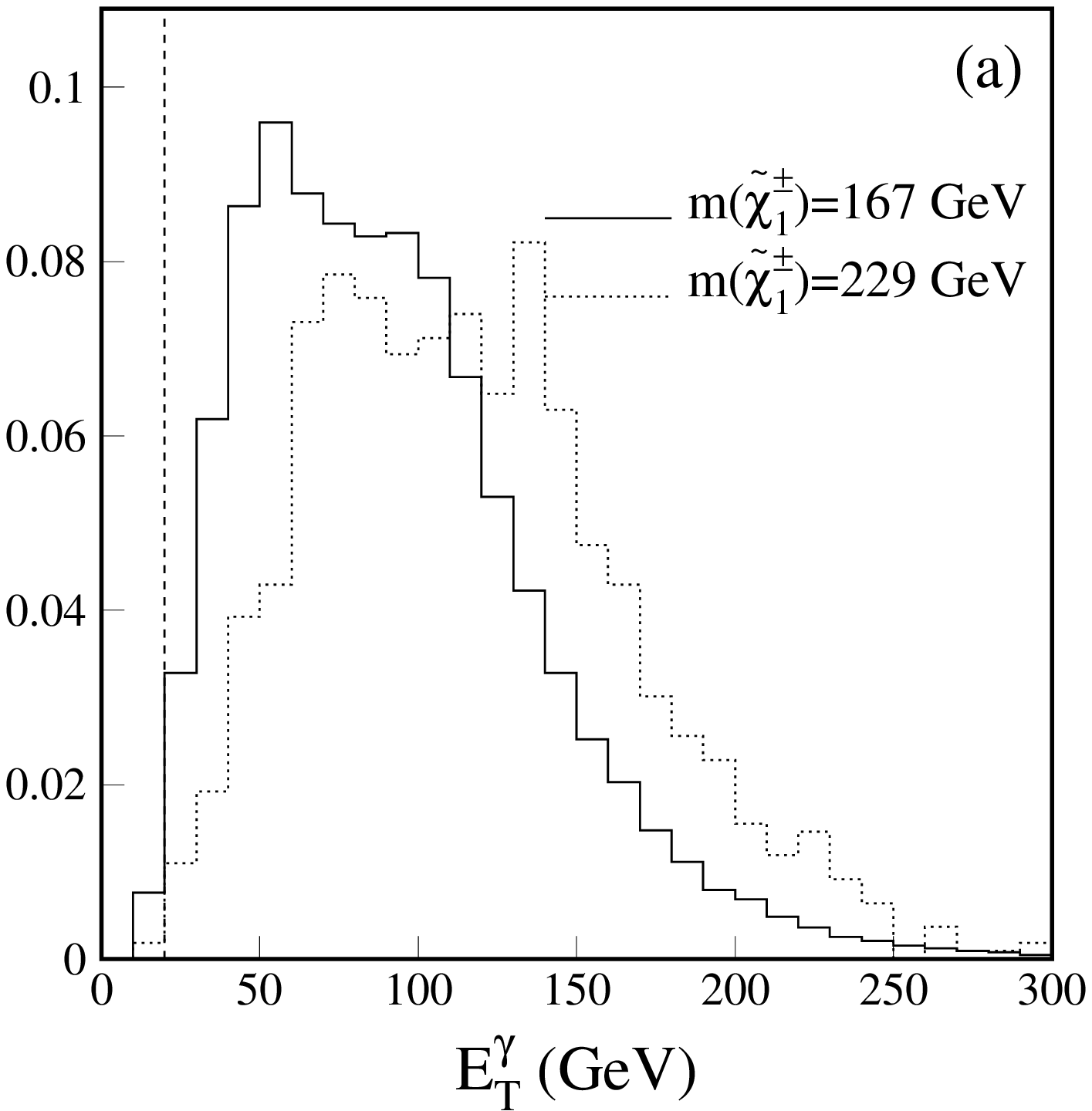}
              \epsfysize=3.0in\epsfbox{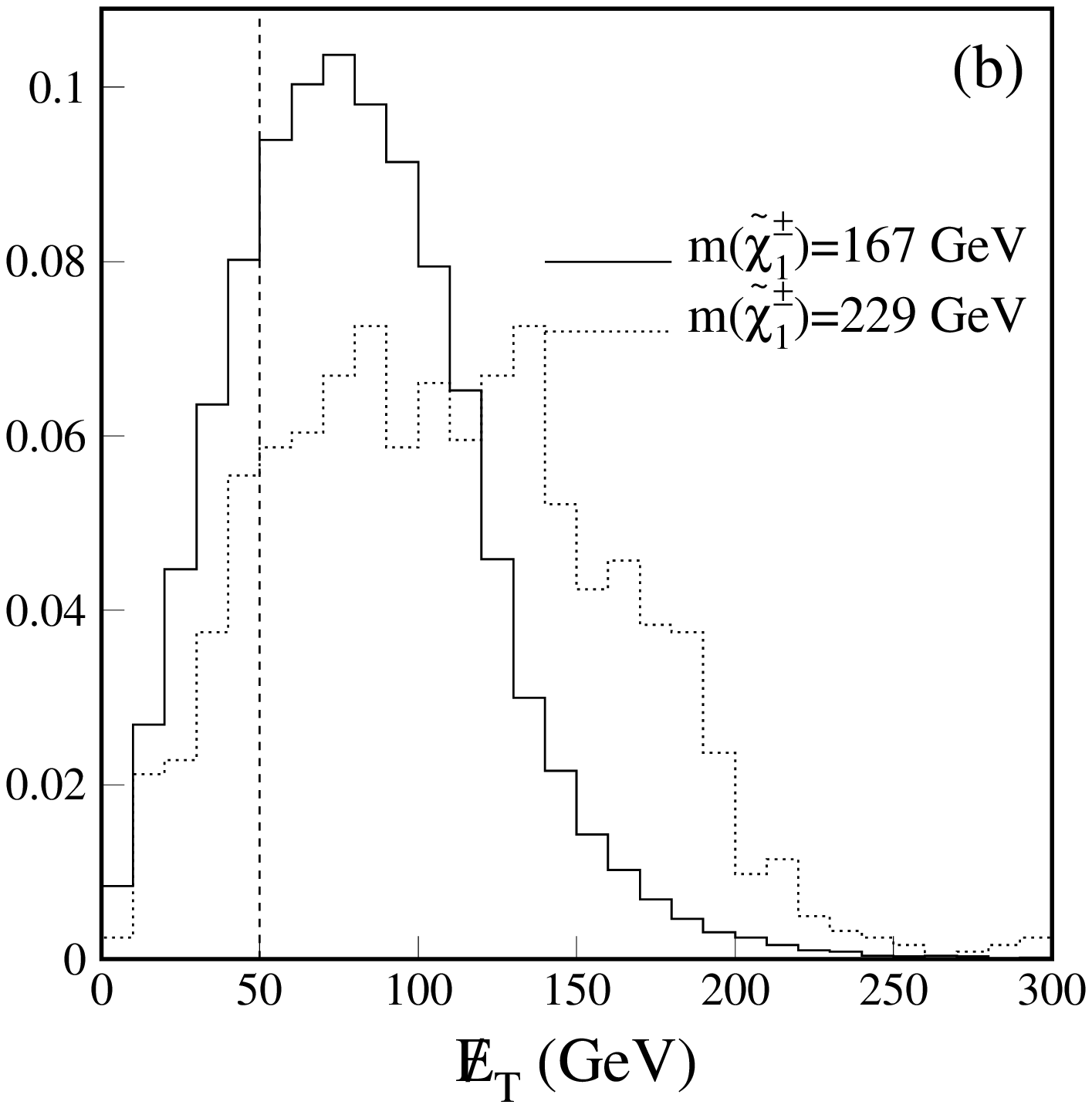}}
  \caption{The photon $E_T$ (a) and event \protect\met\ (b) distributions
           for models with a short-lived higgsino-liked $\tilde\chi^0_1$ 
           as the {\sc nlsp} with $m_{\tilde\chi^\pm_1}=167,\ 229$~GeV 
           ($\Lambda=80,\ 110$~TeV). All distributions are normalized to
           have unit area.}
  \label{fig:p5}
\end{figure}

\begin{figure}[htbp]
  \centerline{\epsfysize=3.5in\epsfbox{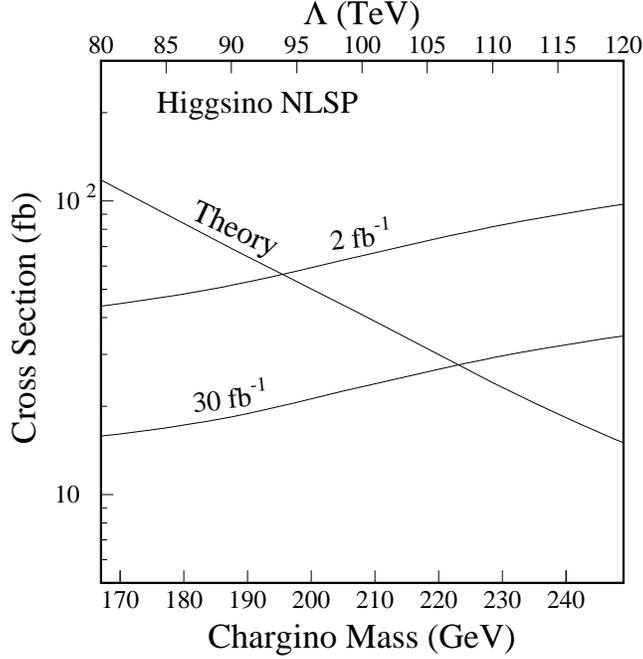}}
  \caption{The $5\sigma$ discovery cross section curves as functions of the 
           supersymmetry breaking scale $\Lambda$ and the lighter chargino 
           mass for the model line 4 along with the theoretical cross sections. 
           The $\tilde h$ is assumed to be short-lived.}
  \label{fig:p5lim}
\end{figure}

For most of the parameter space, $\tilde\chi^0_1$ will predominantly 
decay to $\gamma\tilde G$ if it is the {\sc nlsp}. However if \nlsp\ is
higgsino-like ($\tilde h$), $\tilde\chi^0_1\to Z\tilde G$ and 
$\tilde\chi^0_1\to h\tilde G$ decays could have significant branching ratios 
in some regions of the parameter space of non-minimal GMSB models. For the 
Run II studies, these models are defined to have fixed values of
$$N=2,\ \ \frac{M_m}{\Lambda}=3,\ \ \tan\beta=3,\ \ 
 \frac{\mu}{M_1}=-0.75$$
with $\Lambda$ varying. Here $M_1$ is the mass parameter associated with 
the $U(1)_Y$ symmetry.
In these models, the lightest neutral higgs boson $h$ 
has a mass around 104~GeV. Pair production of supersymmetric particles may 
result in 
$\tilde\chi^0_1\tilde\chi^0_1\to\gamma\tilde G\ h\tilde G\to 
\gamma b\bar{b}\tilde G\tilde G$ final state which would appear as 
\gbbmet\ events in the detector assuming prompt $\tilde\chi^0_1$ decays. 
These events are characterized by high $E_T$ photons and large \met\ as 
shown in the Fig.~\ref{fig:p5} for $\Lambda=80, 110$~TeV.
They can be selected using the \gbjmet\ selection criteria discussed in 
Section~\ref{sec:gbjmet}. The detection efficiencies and $N_s/\delta N_b$
significances are shown in Table~\ref{tab:p5} for different values of
$\Lambda$. Most of the events selected are due to the $\gamma h$ production
with $h\to b\bar{b}$. However, a non-negligible fraction of the events 
is actually due to the $\gamma Z+X$ with $Z\to b\bar{b}$. The discovery reach 
in $\Lambda$ and $m_{\tilde\chi^\pm_1}$ is shown in Fig.~\ref{fig:p5lim}
for \ldt=2, 30~fb$^{-1}$.

\begin{table}[htbp]
  \begin{tabular}{c|ccccc}
     $\Lambda$ (TeV)              &  80 &  90 & 100 & 110 & 120 \\ \hline
     $\sigma_{th}$ (fb)           & 118 &  69 &  41 & 24  & 15 \\
     $m_{\tilde\chi^\pm_1}$ (GeV) & 167 & 188 & 208 & 229 & 249 \\
     $m_{\tilde\chi^0_1}$ (GeV)   & 154 & 176 & 197 & 218 & 239 \\
     $m_h$ (GeV)                  & 103 & 104 & 105 & 105 & 106 \\
     ${\rm Br}(\tilde\chi^0_1\to\gamma\tilde G)$ & 0.38 & 0.21 & 0.15 & 0.10
  & 0.08 \\
     ${\rm Br}(\tilde\chi^0_1\to h\tilde G)$  & 0.38 & 0.52 & 0.59 & 0.63  
& 0.66 \\ \hline
     $\epsilon$ (\%)               &  8.0 & 6.8 & 5.4 & 4.3 & 3.6 \\
     \protect\rsb\ (2 fb$^{-1}$)   &  13  & 6.7 & 3.2 & 1.5 & 0.8 \\ 
     \protect\rsb\ (30 fb$^{-1}$)  &  38  &  19 & 8.9 & 4.1 & 2.2 \\ 
  \end{tabular}
  \caption{The theoretical cross section, $\tilde\chi^\pm_1$ and $h$ masses, 
           \protect\nlsp\ decay branching ratios, detection efficiency of
           the \protect\gbjmet\ selection, and 
           significances for different values of $\Lambda$ for
           the models with a short-lived higgsino as the {\sc nlsp}.
           The relative statistical error on the efficiency is typically
           4\%. The background cross section is assumed to be 0.9~fb with 
           a 20\% systematic error.}
  \label{tab:p5}
\end{table}

\begin{figure}[htbp]
  \centerline{\epsfysize=3.5in\epsfbox{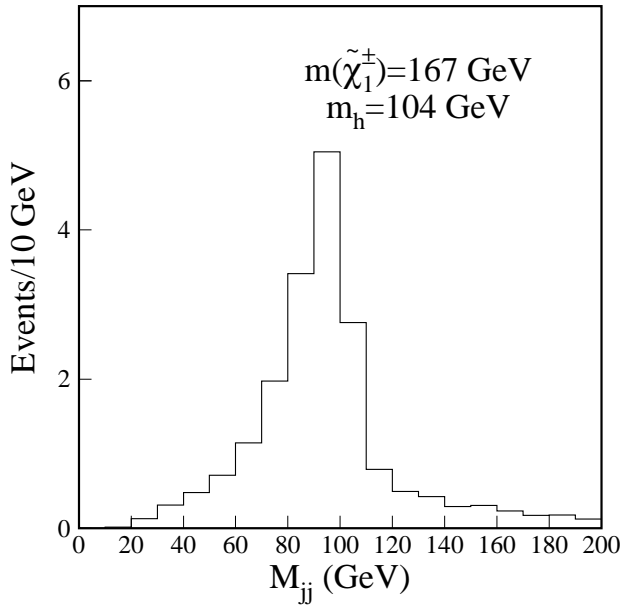}}
  \caption{The invariant mass distribution of the two leading jets for 
           $m_{\tilde\chi^\pm_1}=167$~GeV ($\Lambda=80$~TeV) and 
           \protect\ldt=2~fb$^{-1}$. Note that 19 signal and less than 
           2 background events are expected.}
  \label{fig:p5mjj}
\end{figure}

If an excess is seen in the \gbjmet\ final state, it will be of great interest
to reconstruct the di-jet invariant mass. Figure~\ref{fig:p5mjj} shows the 
invariant mass distribution of the two leading jets for $\Lambda=80$~TeV 
and \ldt=2~fb$^{-1}$. A mass peak around $m_h=104$~GeV is clearly
identifiable. The asymmetry in the mass distribution is partly due to 
$\gamma Z+X$ contribution and partly due to the effect of gluon radiation
and the energy outside the jet cone.

\section{Summary}
In this paper, observable background cross sections for the six final states 
in which new physics might manifest itself are estimated. All the final 
states studied
are found to have small backgrounds. Implications of the analyses of these
final states for future Tevatron runs are discussed in the framework of Gauge 
Mediated Supersymmetry Breaking models. Potential discovery reaches in 
supersymmetry parameter space for integrated luminosities of 2 and 
30~fb$^{-1}$ are examined for models with different {\sc nlsp}. 
Though the selection criteria are not optimized for the models discussed
and not all final states are investigated, the study does show that the 
upgraded D\O\ experiment at the improved Tevatron collider has great
potentials for discovery.

\section{Acknowledgement}
The author would like to thank D.~Cutts, K.~Del~Signore, G.~Landsberg, 
S.~Martin, H.~Montgomery, S.~Thomas, D.~Toback, A.~Turcot, H.~Weerts, 
and J.~Womersley for their assistance in the course of this study and/or 
their critical reading of this writeup and X. Tata for pointing out a 
mistake in one of the background estimations.

\end{document}